\documentclass[12pt]{article}
\pdfoutput=1
\usepackage[colorlinks,linkcolor=Blue,citecolor=Blue,bookmarks,bookmarksnumbered]{hyperref}
\usepackage[scaled=0.85]{helvet}
\usepackage{amsmath,amssymb,accents,mathrsfs,XoohmE}
\usepackage{graphicx,color}
\usepackage{booktabs}
\usepackage{multirow}
\usepackage{placeins}
\usepackage{amsmath}

\usepackage{XoohmE}

\definecolor{Green}  {rgb}{0.10,0.70,0.10} 
\definecolor{Orange} {rgb}{1.00,0.50,0.15} 
\definecolor{Red}    {rgb}{0.90,0.00,0.12} 
\definecolor{Purple} {rgb}{0.50,0.25,0.55} 
\definecolor{Turque} {rgb}{0.00,0.65,0.85} 
\definecolor{Blue}   {rgb}{0.00,0.00,1.00} 
\definecolor{Magenta}{rgb}{1.00,0.00,1.00} 
\definecolor{Gold}   {rgb}{1.00,0.75,0.25} 
\definecolor{Seaweed}{rgb}{0.01,0.24,0.09} 
\definecolor{Brown}  {rgb}{0.43,0.26,0.32} 
\definecolor{grey1}  {rgb}{0.20,0.20,0.20} 
\definecolor{grey2}  {rgb}{0.40,0.40,0.40} 
\definecolor{grey3}  {rgb}{0.60,0.60,0.60} 
\definecolor{grey4}  {rgb}{0.80,0.80,0.80} 
\definecolor{grey5}  {rgb}{0.90,0.90,0.90} 
\def\C#1#2{{\ifcase#1\or
             \color{Green}\or \color{Orange}\or \color{Red}\or
              \color{Purple}\or \color{Turque}\or \color{Blue}\or
               \color{Magenta}\or \color{Gold}\or \color{Seaweed}\or
                \color{Brown}\or\color{grey1}\or\color{grey2}\or
                 \color{grey3}\else\color{grey4}\fi#2}}

\definecolor{Slate} {rgb}{0.00,0.45,0.55}

\def\fracm#1#2{\hbox{\large{${\frac{{#1}}{{#2}}}$}}}

\def\be{\begin{equation}}
\def\ee{\end{equation}}
\newcommand{\bea}{\begin{eqnarray}}
\newcommand{\eea}{\end{eqnarray}}
\newcommand{\ena}{\end{eqnarray}}

\def\pp{{\mathchoice
          {
              \kern 1pt%
              \raise 1pt
              \vbox{\hrule width5pt height0.4pt depth0pt
                    \kern -2pt
                    \hbox{\kern 2.3pt
                          \vrule width0.4pt height6pt depth0pt
                          }
                    \kern -2pt
                    \hrule width5pt height0.4pt depth0pt}%
                    \kern 1pt
           }
            {
              \kern 1pt%
              \raise 1pt
              \vbox{\hrule width4.3pt height0.4pt depth0pt
                    \kern -1.8pt
                    \hbox{\kern 1.95pt
                          \vrule width0.4pt height5.4pt depth0pt
                          }
                    \kern -1.8pt
                    \hrule width4.3pt height0.4pt depth0pt}%
                    \kern 1pt
            }
            {
              \kern 0.5pt%
              \raise 1pt
              \vbox{\hrule width4.0pt height0.3pt depth0pt
                    \kern -1.9pt  
                    \hbox{\kern 1.85pt
                          \vrule width0.3pt height5.7pt depth0pt
                          }
                    \kern -1.9pt
                    \hrule width4.0pt height0.3pt depth0pt}%
                    \kern 0.5pt
            }
            {
              \kern 0.5pt%
              \raise 1pt
              \vbox{\hrule width3.6pt height0.3pt depth0pt
                    \kern -1.5pt
                    \hbox{\kern 1.65pt
                          \vrule width0.3pt height4.5pt depth0pt
                          }
                    \kern -1.5pt
                    \hrule width3.6pt height0.3pt depth0pt}%
                    \kern 0.5pt
            }
        }}

\def\mm{{\mathchoice
                       {
                             \kern 1pt
               \raise 1pt    \vbox{\hrule width5pt height0.4pt depth0pt
                                  \kern 2pt
                                  \hrule width5pt height0.4pt depth0pt}
                             \kern 1pt}
                       {
                            \kern 1pt
               \raise 1pt \vbox{\hrule width4.3pt height0.4pt depth0pt
                                  \kern 1.8pt
                                  \hrule width4.3pt height0.4pt depth0pt}
                             \kern 1pt}
                       {
                            \kern 0.5pt
               \raise 1pt
                            \vbox{\hrule width4.0pt height0.3pt depth0pt
                                  \kern 1.9pt
                                  \hrule width4.0pt height0.3pt depth0pt}
                            \kern 1pt}
                       {
                           \kern 0.5pt
             \raise 1pt  \vbox{\hrule width3.6pt height0.3pt depth0pt
                                  \kern 1.5pt
                                  \hrule width3.6pt height0.3pt depth0pt}
                           \kern 0.5pt}
                       }}

\def\ad{{\kern0.5pt
                   \alpha \kern-5.05pt \raise5.8pt\hbox{$\textstyle.$}\kern
0.5pt}}

\def\bd{{\kern0.5pt
                   \beta \kern-5.05pt \raise5.8pt\hbox{$\textstyle.$}\kern
0.5pt}}

\def\qd{{\kern0.5pt
                   q \kern-5.05pt \raise5.8pt\hbox{$\textstyle.$}\kern
0.5pt}}
\def\Dot#1{{\kern0.5pt
     {#1} \kern-5.05pt \raise5.8pt\hbox{$\textstyle.$}\kern
0.5pt}}

\catcode`@=11
\def\un#1{\relax\ifmmode\@@underline#1\else
        $\@@underline{\hbox{#1}}$\relax\fi}
\catcode`@=12

\def\a{\alpha}
\def\b{\beta}
\def\c{\chi}
\def\d{\delta}
\def\e{\epsilon}

\def\g{\gamma}

\def\i{\iota}
\def\j{\psi}
\def\k{\kappa}
\def\l{\lambda}

\def\o{\omega}

\def\q{\theta}
\def\r{\rho}
\def\s{\sigma}
\def\t{\tau}

\def\D{\Delta}

\def\L{\Lambda}
\def\O{\Omega}
\def\P{\Pi}

\def\S{\Sigma}

\def\ca{{\cal A}}

\def\cf{{\cal F}}

\def\dslash{\not{\hbox{\kern-2pt $\partial$}}}
\def\Dslash{\not{\hbox{\kern-4pt $D$}}}
\def\pslash{\not{\hbox{\kern-2.3pt $p$}}}
 \newtoks\slashfraction
 \slashfraction={.13}
 \def\slash#1{\setbox0\hbox{$ #1 $}
 \setbox0\hbox to \the\slashfraction\wd0{\hss \box0}/\box0 }
 
\def\kcr{{\hbox{\ro \char'170}}}                
\def\ktl{{\hbox{\ro \char'170}}}        
\def\ktr{{\hbox{\ro \char'170}}}        
\def\kbl{{\hbox{\ro \char'170}}}        
\def\kbr{{\hbox{\ro \char'170}}}        

\def\plpl{\raise-2pt\hbox{$\raise3pt\hbox{$_+$}\hskip-6.67pt\raise0.0pt
\hbox{$^+$}\hskip 0.01pt$}}
\def\mimi{\raise-2pt\hbox{$\raise3pt\hbox{$_-$}\hskip-6.67pt\raise0.0pt
\hbox{$^-$}\hskip 0.01pt$}} 

\def\bo{{\raise.15ex\hbox{\large$\Box$}}}               
\def\pa{\partial}                                       
\def\TH{{\raise.2ex\hbox{$\displaystyle \bigodot$}\mskip-4.7mu \llap H \;}}
\def\face{{\raise.2ex\hbox{$\displaystyle \bigodot$}\mskip-2.2mu \llap {$\ddot
        \smile$}}}                                      

\def\dt#1{\on{\hbox{\bf .}}{#1}}                
\def\Dot#1{\dt{#1}}

\def\Hat#1{\widehat{#1}}                        
\def\Bar#1{\overline{#1}}                       
\def\leftrightarrowfill{$\mathsurround=0pt \mathord\leftarrow \mkern-6mu
        \cleaders\hbox{$\mkern-2mu \mathord- \mkern-2mu$}\hfill
        \mkern-6mu \mathord\rightarrow$}
\def\dvec#1{\vbox{\ialign{##\crcr
        \leftrightarrowfill\crcr\noalign{\kern-1pt\nointerlineskip}
        $\hfil\displaystyle{#1}\hfil$\crcr}}}           
\def\dt#1{{\buildrel {\hbox{\LARGE .}} \over {#1}}}     

\def\fracm#1#2{\hbox{\large{${\frac{{#1}}{{#2}}}$}}}
\def\sfrac#1#2{{\vphantom1\smash{\lower.5ex\hbox{\small$#1$}}\over
        \vphantom1\smash{\raise.4ex\hbox{\small$#2$}}}} 
\def\bfrac#1#2{{\vphantom1\smash{\lower.5ex\hbox{$#1$}}\over
        \vphantom1\smash{\raise.3ex\hbox{$#2$}}}}       
\def\afrac#1#2{{\vphantom1\smash{\lower.5ex\hbox{$#1$}}\over#2}}    

\def\pa{\partial}      
\let\bm\relax
\newcommand{\bm}[1]{{\boldsymbol{#1}}}

\def\ad{{\dot{\alpha}}}
\def\bd{{\dot{\beta}}}

 \font\rOpe=cmsy10                        
 \def\ktl{{\hbox{\rOpe\char'170}}}        
 \def\kbl{{\hbox{\rOpe\char'170}}}        
 \def\kcr{{\reflectbox{\rOpe\char'170}}}        
 \def\ktr{{\reflectbox{\rOpe\char'170}}}        
 \def\kbr{{\reflectbox{\rOpe\char'170}}}        
 \def\Border{\vbox{\hsize0pt
        \setlength{\unitlength}{1mm}
        \newcount\xco
        \newcount\yco
        \xco=-21
        \yco=12
        \begin{picture}(0,0)(-7.5,0)
        \put(\xco,\yco){$\ktl$}
        \advance\yco by-1
        {\loop
        \put(\xco,\yco){$\kcr$}
        \advance\yco by-2
        \ifnum\yco>-240
        \repeat
        \put(\xco,\yco){$\kbl$}}
        \xco=170
        \yco=12
        \put(\xco,\yco){$\ktr$}
        \advance\yco by-1
        {\loop
        \put(\xco,\yco){$\kcr$}
        \advance\yco by-2
        \ifnum\yco>-240
        \repeat
        \put(\xco,\yco){$\kbr$}}
        \put(-19.5,13){\scalebox{.6065}{%
         University of Maryland Center for String and Particle  Theory \&\ Physics Department%
        |University of Maryland Center for String and Particle  Theory \&\ Physics Department}}
        \put(-19.5,-241.5){\scalebox{.5835}{%
         ****University of Maryland * Center for String and
         Particle  Theory* Physics Department****University of Maryland *Center
        for String and Particle  Theory* Physics Department}}
        \end{picture}
        \par\vskip-8mm}}
\definecolor{UMred}{rgb}{.9,.05,.2}
\definecolor{HUblue}{rgb}{.0,.3,.7}

\definecolor{Red}    {rgb}{0.90,0.00,0.12} 
\definecolor{Blue}   {rgb}{0.00,0.00,1.00} 
\definecolor{Green}  {rgb}{0.10,0.70,0.10} 
\definecolor{Turque} {rgb}{0.00,0.65,0.85} 
\definecolor{Orange} {rgb}{1.00,0.50,0.15} 
\definecolor{Magenta}{rgb}{1.00,0.00,1.00} 
\definecolor{Gold}   {rgb}{1.00,0.75,0.25} 
\definecolor{Seaweed}{rgb}{0.01,0.24,0.09} 
\definecolor{Purple} {rgb}{0.50,0.25,0.55} 
\definecolor{Brown}  {rgb}{0.43,0.26,0.32} 
\definecolor{grey1}  {rgb}{0.20,0.20,0.20} 
\definecolor{grey2}  {rgb}{0.40,0.40,0.40} 
\definecolor{grey3}  {rgb}{0.60,0.60,0.60} 
\definecolor{grey4}  {rgb}{0.80,0.80,0.80} 
\definecolor{grey5}  {rgb}{0.90,0.90,0.90} 
\def\C#1#2{{\ifcase#1\or
             \color{Red}\or \color{Green}\or \color{Blue}\or\
              \color{Turque}\or \color{Orange}\or \color{Magenta}\or 
               \color{Gold}\or \color{Seaweed}\or \color{Purple}\or
                \color{Brown}\or\color{grey1}\or\color{grey2}\or
                 \color{grey3}\else\color{grey4}\fi#2}}

\definecolor{Slate} {rgb}{0.00,0.45,0.55}

\newdimen\parshift\parshift=\parindent
\catcode`@=11
 \long\def\@footnotetext#1{\insert\footins{\reset@font\footnotesize
           \interlinepenalty\interfootnotelinepenalty\splittopskip%
            \footnotesep\splitmaxdepth\dp\strutbox\floatingpenalty\@MM%
             \hsize\columnwidth\addtolength{\hsize}{-2\parindent}
              \@parboxrestore\protected@edef\@currentlabel%
              {\csname p@footnote\endcsname\@thefnmark}%
                \color@begingroup%
                 \@makefntext{\rule\z@\footnotesep\ignorespaces#1%
                  \@finalstrut\strutbox}%
                \color@endgroup}}
 \long\def\@makefntext#1{\hglue\parshift%
           \vbox{\noindent\baselineskip=11pt plus.5pt minus.5pt\hb@xt@0em{\hss\@makefnmark\kern1pt}#1}}
\catcode`@=12

\newskip\humongous \humongous=0pt plus 1000pt minus 1000pt
\def\caja{\mathsurround=0pt}
\def\eqalign#1{\,\vcenter{\openup2\jot \caja
        \ialign{\strut \hfil$\displaystyle{##}$&$
        \displaystyle{{}##}$\hfil\crcr#1\crcr}}\,}
\newif\ifdtup
\def\panorama{\global\dtuptrue \openup2\jot \caja
        \everycr{\noalign{\ifdtup \global\dtupfalse
        \vskip-\lineskiplimit \vskip\normallineskiplimit
        \else \penalty\interdisplaylinepenalty \fi}}}

\makeatletter
\def\section{\@startsection{section}{1}{\z@}
        {3ex plus-1ex minus-.2ex}{1pt plus1pt}{\large\sf\bfseries\boldmath}}
\def\subsection{\@startsection{subsection}{2}{\z@}
         {1.5ex plus-1ex minus-.2ex}{0.01pt plus1pt}{\sf\slshape}}
\def\subsubsection{\@startsection{subsubsection}{3}{\z@}
          {1.5ex plus-1ex minus-.2ex}{0.01pt plus0.2pt}{\sf\boldmath}}
\def\paragraph{\@startsection{paragraph}{4}{\z@}
           {.75ex \@plus.5ex \@minus.2ex}{-2mm}{\sf\bfseries\boldmath}}
\makeatother

 \allowdisplaybreaks
 \seceq

\begin{document}

\thispagestyle{empty}
%
\noindent{\small
\hfill{HET-1783}  \\ 
$~~~~~~~~~~~~~~~~~~~~~~~~~~~~~~~~~~~~~~~~~~~~~~~~~~~~~~~~~~~~~~~~~$
$~~~~~~~~~~~~~~~~~~~~~~~~~~~~~~~~~~~~~~~~~~~~~~~~~~~~~~~~~~~~~~~~~$
{}
}
\vspace*{8mm}
\begin{center}
{\large \bf
On Linearized Nordstr\" om Supergravity in \vskip0.1pt
Eleven and Ten Dimensional Superspaces (2)}   \\   [12mm]
{\large {
S.\ James Gates, Jr.,\footnote{sylvester$_-$gates@brown.edu}$^{a}$}
Yangrui Hu\footnote{yangrui$_-$hu@brown.edu}$^{a}$,
Hanzhi Jiang\footnote{hanzhi$_-$jiang@alumni.brown.edu}$^{a,b}$, 
and S.-N. Hazel Mak\footnote{sze$_-$ning$_-$mak@brown.edu}$^{a}$
}
\\*[12mm]
\emph{
\centering
$^{a}$Brown Theoretical Physics Center, and 
\\[1pt]
Department of Physics, Brown University,
\\[1pt]
Box 1843, 182 Hope Street, Barus \& Holley 545,
Providence, RI 02912, USA 
\\[12pt]
${}^{b}$Department of Physics \& Astronomy,
\\[1pt]
Rutgers University,
 Piscataway, NJ 08855-0849, USA
}
 \\*[78mm]
{ ABSTRACT}\\[4mm]
\parbox{142mm}{\parindent=2pc\indent\baselineskip=14pt plus1pt
We present aspects of the component description of linearized Nordstr\" om 
Supergravity in eleven and ten dimensions.  The presentation includes low 
order component fields in the supermultiplet, the supersymmetry variations 
of the scalar graviton and gravitino trace, their supercovariantized field strengths, 
and the supersymmetry commutator algebra of these theories. }
 \end{center}
\vfill
\vfill
\clearpage
%

\newpage
\section{Introduction}

A mathematically consistent (however requiring restrictions on the allowed general
coordinate transformations) and simplified version of gravitation is provided by a
variant \cite{N1,N2} that may be called ``Nordstr\" om Gravity.''  In a previous work
\cite{NordSG1}, we initiated a program of investigating\footnote{In this previous 
work, a substantial citation review is undertaken and interested readers are encouraged 
to familiarize themselves with the literature via this means.} whether one can construct 
Nordstr\" om Supergravity extensions in eleven and ten 
dimensional spacetimes of this simplified of limit of gravitation at the linearized level.

One pointed motivation for our efforts has been the recent progress  \cite{PF1,PF2}
 in the derivation of M-Theory corrections to 11D Supergravity.  A series of procedures 
 connecting the corrections to a 3D, $\cal N$ = 2 Chern-Simons theory \cite{CST1,CST2,CST3}
 (used in a role roughly analogous to world-sheet $\s$-model $\b$-function calculations
 for string corrections) has been successfully demonstrated.  Though the works in 
 \cite{PF1,PF2} have presented a method of deriving these corrections beyond the 
supergravity limit, these {\it {solely}} treat purely bosonic M-Theory corrections, with 
no equivalent results describing fermionic corrections.  One traditional way of 
accomplishing this is to embed the purely bosonic results into a superspace formulation.  
This impels us to a renewed interest in 11D supergravity in superspace.

The goal we are pursuing is to find a Salam-Strathdee superspace 
\cite{Ssp8c}, as modified by Wess \& Zumino \cite{WZ1,WZ2}, such that superspace Bianchi 
identities do {\em {not imply}} equations of motion for the component fields contained within 
the superspace description of Nordstr\" om SG. In particular, we are {\em {not}} currently 
investigating the prospect of writing action formulae for such supermultiplets of fields.  
While actions are the ``gold standard,'' it is useful to recall (as done below), this is not
the first time the off-shell structure of a supergravity theory has been explored, {\it {without}}
the additional exploration of an action principle.

This distinguishes our work from efforts taken by other.  For example, there is a substantial 
literature that uses the concept of ``pure spinors'' \cite{P1,P2,P3,P4,P5,P6,P7} where the 
endpoint of action principles (see especially \cite{P6,P7}) has been presented.  While 
classically such approaches appear to work, there are troubling signs \cite{Q1,Q2,Q3,Q4}
that more needs to be done to justify complete acceptance at the level of quantum
theory.  

Perhaps an intuitive way to argue is in order to achieve quantization, it should be 
implemented in terms of variables that are basically free. The non-minimal pure spinor 
goes some way to giving a free resolution, but the resulting space of fields does not 
have a well-defined trace. The work of \cite{Q3} offered a ``fix'' but {\em {if}} in the end 
the prescription still is not an integral over free variables with no other qualifications, 
then a proper quantization is still likely to be impeded.  

Therefore, we adopt the rather more cautious approach by raising the query of whether
it is possible to follow the pathway established by Wess \& Zumino \cite{WZ1,WZ2} wherein
a ``simple'' Salam-Strathdee superspace is used as a starting point to building, in our
case, Nordstr\" om SG in which Bianchi identities do not force equations of motion.

While it is often overlooked, the first off-shell description of 4D, $\cal N$ = 1 supergravity
was actually carried out by Breitenlohner \cite{B1} who took an approach equivalent to starting
with the component fields of the Wess-Zumino gauge 4D, $\cal N$ = 1 vector supermultiplet 
$(v{}_{\un{a}}, \, \l_b, {\rm d})$ together with their familiar SUSY transformation laws,
\be \eqalign{
{\rm D}_a \, v{}_{\un b} ~&=~  (\g_{\un b}){}_a {}^c \,  \l_c  ~~~, ~~~ {~~~ 
~~~~~~~~~~~~~~~~~~~~~~~~~~~~~~~~} \cr
{\rm D}_a \l_b ~&=~   - \,i \, \fracm 14 ( [\, \g^{\un c}\, , \,  \g^{\un d} 
\,]){}_a{}_b \, (\,  \pa_{\un c}  \, v{}_{\un d}    ~-~  \pa_{\un d} \, v{}_{\un c}  \, )
~+~  (\g^5){}_{a \,b} \,    {\rm d} ~~, {~~~~~~} \cr
{\rm D}_a \, {\rm d} ~&=~  i \, (\g^5 \g^{\un c} ){}_a {}^b \, 
\,  \pa_{\un c} \l_b  ~~~, \cr
} \label{V1}
\ee
followed by choosing as the gauge group the space time translations, SUSY generators, and the
spin angular momentum generators as well as allowing additional internal symmetries.  For the 
space time translations, this requires a series of replacements of the fields according to:
\be  \eqalign{
 v{}_{\un b} ~& \to ~  h{}_{\un{b} \, \un{c}} ~~~,   ~~~
  \l_b ~ \to ~ \psi{}_{\un{c} \, b}  \,~~~,   ~~~
 {\rm d} ~ \to ~ A{}_{\un{c}}  ~~~~~, 
}  \label{V2} \ee
(in the notation in \cite{B1} $A{}_{\un{a}} $ = $B^5{}_{\un{a}}$)
while for the SUSY generators, the replacements occur according to:
 \be  \eqalign{
 v{}_{\un b} ~& \to ~  \chi{}_{\un{b} \, c} ~~~,   ~~~
  \l_b ~ \to ~ \phi{}_{b \, c}  \, ~~~,   ~~~
 {\rm d} ~ \to ~ \chi{}_{c}{}^5  ~~~~~, 
} \ee
and finally for the spin angular momentum generator, a replacement of 
 \be  \eqalign{
 v{}_{\un b} ~& \to ~  \o{}_{\un{b} \, \un{c} \, \un{d}} ~~~,  ~~~
  \l_b ~ \to ~ \chi{}_{b \, \un{c} \, \un{d}}  \, ~~~,   ~~~
 {\rm d} ~ \to ~ D{}_{\un{c} \, \un{d}}  ~~~~~, 
} \ee
was used.  However, to be more exact, Breitenlohner also allowed for more
symmetries like chirality to be included.  Because the vector supermultiplet
was off-shell (up to WZ gauge transformations) the resulting supergravity theory
was off-shell and included a redundant set of auxiliary component fields,  i.\ e.\ this is
not an irreducible description of supergravity.  But as seen from (\ref{V2}) the
supergravity fields were all present and together with the remaining component fields
a complete superspace geometry can be constructed.

In our approach to Nordstr\" om SG, the analog of the Wess-Zumino gauge
4D, $\cal N$ = 1 vector supermultiplet is played by a scalar superfield in any
of the 11D or 10D superspaces to be studied.  This scalar superfield guarantees
off-shell supersymmetry.  However, like the approach taken by Breitenlohner,
the resulting theory is expected to be reducible.  Also like this earlier approach,
the question of an action principle is not addressed.

The structure of the remainder of this work looks as follows.

In chapter two, a review of 4D, $\cal N$ = 1 supergravity in superspace is given.  
This proves a detailed description of how to extract component results from the
superspace geometry.  Using the foundation in  {\it {Superspace}} \cite{SpRSp8BK}, 
the general formalism for obtaining component level results is reviewed.  In
this context the composition rules for the parameter of spacetime translations,
parameters of SUSY transformations, and Lorentz transformations are presented
relating these to supergeometrical quantities is given.
Next the SUSY transformation rules for the frame field, gravitino field, and
spin connection relating these to supergeometrical quantities are given.
Finally, the ``supercovariantized'' field strengths of the frame field, gravitino field, 
and spin connection relating these to supergeometrical quantities are given.
and related to supergeometrical quantities.  This chapter ends with the linearization
of these results.

Chapter three uses the technology of the second chapter to present component
level of the Nordstr\" om SG for 11D, $\cal N $ = 1 superspace, 10D, $\cal N $ = 2A 
superspace,  10D, $\cal N $ = 2B superspace, and  10D, $\cal N $ = 1 superspace.
Component level descriptions of the local SUSY commutator algebras are provided.
Linearized curvatures and torsions supertensors are presented and the supersymmetry
variations of the linearized Nordstr\" om ``scalar'' graviton, the linearized
spin-1/2 Nordstr\" om gravitino, and the spin connection are obtained.

Chapter four is a short chapter in comparison to the two that precede it.  In 4D,
$\cal N$ = 1 supergravity \cite{S1,SFSG}, the concept of the ``chiral compensator''
was introduced some time ago.  We demonstrate evidence that such a compensator
exist for the 10D, $\cal N $ = 2B superspace.  This is unique among supergravity
theories in eleven and ten dimensions.  However, we also present evidence that
though such a chiral superfields appear to consistently exist, the linearized
Nordstr\"om superspace is such that a chiral superfield of this type cannot be
used as a compensator.

The fifth chapter is devoted to our conclusions and a summary.

\newpage
\section{Superspace Perspective On Component Results}

In our previous paper \cite{NordSG1}, we restricted our focus solely to superfield 
considerations in eleven and ten dimensions.  However given those result, the technology 
developed in {\em {Superspace}} \cite{SpRSp8BK} allows a presentation of some of 
the component results.  In particular, the equations indicated in section (5.6) in this book 
can be applied to the case of eleven and ten dimensions.  This is true even though the 
sole focus of the book is the case of 4D, $\cal N$ = 1 supersymmetry.  Nonetheless, the 
discussion in the book can be easily modified for use in 11D and 10D 
superspace theories.  The relevant equations were designated as (5.6.13), (5.6.16) - 
(5.6.18), (5.6.21), (5.6.22) - (5.6.24), (5.6.28), (5.6.33), and (5.6.34).  For the convenience 
of the reader, we bring these results all together in the text to follow.  After this
chapter, these are all going to be appropriately modified for the cases 
of 11D, $\cal N$ = 1, 10D, $\cal N$ = 2A, 10D, $\cal N$ = 2B, and 10D, $\cal N$ = 1 
superspaces, respectively.
\subsection{Recollection of 4D, $\cal N$ = 1 Component/Superspace Results}

In the context of 4D, $\cal N$ = 1 superspace supergravity, we may distinguish 
among three types of symmetries: \vskip1mm
$~$ \noindent
(a.) space time translations with generator $iK_{GC}(\xi^{\un {m}})$,
dependent \newline \indent $~~~~~~~$
on local parameters $\xi^{\un {m}}(x)$,
$~$ \newline \indent $~$
(b.) SUSY transformations with generator $iK_{Q}(\e^{\un {\a}})$ dependent on 
\newline \indent $~~~~~~~$
local parameters $\e^{\un {\a}}(x)$, and
$~$ \newline \indent $~$
(c.) tangent space transformations with generator $iK_{TS}(\l^{\iota})$ depend 
\newline \indent $~~~~~~~$
-ent on local parameters $\l^{\iota}(x)$.
\vskip0.9mm \noindent
The tangent space transformations act as ``internal angular momentum,'' chirality, etc. on
all ``flat indices'' associated with the superspace quantities.

The commutator algebra of two SUSY transformations generated by $iK_{Q}(\e{}_1{}^{\un 
{\a}})$, and $iK_{Q}(\e{}_2{}^{\un {\a}})$, respectively takes the form
\begin{equation}
\big[ \, iK_{Q}(\e{}_1) ~,~ iK_{Q}(\e{}_2) \, \big]
~=~iK_{GC}(\xi^{\un  {m}})+iK_{Q}(\e)+ iK_{TS}(\l^{\iota}) ~~~,
\label{e1}
\end{equation}
where the parameters $\xi^{\un  {m}}$, $\e^{\un  {\d}}$, and $\l^{\iota}$ on 
the RHS of this equation are quadratic in $\e{}_1$ and $\e{}_2$, dependent on linear 
and quadratic terms in the gravitino, and linear terms in the superspace torsions and 
curvature supertensors according to:
\begin{align}
\xi^{\un  {m}} ~=&~ -\Big[(\e_1^{\ \un {\a}}\bar{\e}_2^{\ \dot {\un \b}}+\bar{\e}_1^{\ \dot {\un{\b}}}\e_2^{\ 
\un {\a}})T_{\un {\a} \dot {\un{\b}}}^{\ \ \un {c}} +\e_1^{\ \un  {\a}}\e_2^{\ \un  {\b}}T_{\un  {\a}\un  
{\b}}^{\ \ \un  {c}} + \bar{\e}_1^{\ \dot {\un{\a}}}\bar{\e}_2^{\ \dot {\un{\b}}}T_{\dot {\un{\a}}
\dot {\un{\b}}}^{\ \ \un  {c}}\Big] e_{\un  {c}}^{\ \un  {m}} ~~~,  \label{e2} \\
\e^{\un  {\d}} ~=&~ - \Big[ (\e_1^{\ \un  {\a}}\bar{\e}_2^{\ \dot {\un{\b}}}+\bar{\e}_1^{\ \dot {\un{\b}}}
\e_2^{\ \un  {\a}})(T_{\un  {\a}\dot {\un{\b}}}^{\ \ \un  {\d}} + T_{\un  {\a}\dot {\un{\b}}}^{\ \ \un  
{c}}\psi_{\un  {c}}^{\ \un  {\d}}) + \e_1^{\ \un  {\a}}\e_2^{\ \un  {\b}}(T_{\un  {\a}\un  {\b}}^{\ \ \un 
{\d}} + T_{\un  {\a}\un {\b}}^{\ \ \un  {c}}\psi_{\un  {c}}^{\ \un {\d}})  + \bar{\e}_1^{\ \dot {\un{\a}}}
\bar{\e}_2^{\ \dot {\un{\b}}}(T_{\dot {\un{\a}}\dot {\un{\b}}}^{\ \ \un  {\d}}+T_{\dot {\un{\a}}\dot 
{\un{\b}}}^{\ \ \un {c}}\psi_{\un  {c}}^{\ \un  {\d}}) \Big] ~~~,  \label{e4}  \\
\l^{\iota} ~=&~ - \Big[(\e_1^{\ \un  {\a}}\bar{\e}_2^{\ \dot {\un{\b}}} + \bar{\e}_1^{\ \dot {\un{\b}}}\e_2^{\ 
\un  {\a}})(R_{\un  {\a}\dot {\un{\b}}}^{\ \ \ \iota} + T_{\un  {\a}\dot {\un{\b}}}^{\ \ \un  {c}}\phi_{\un 
{c}}^{\ \iota}) + \e_1^{\ \un  {\a}}\e_2^{\ \un  {\b}}(R_{\un  {\a}\un  {\b}}^{\ \ \ \iota} + T_{\un  {\a}
\un  {\b}}^{\ \ \un  {c}}\phi_{\un  {c}}^{\ \iota})  + \bar{\e}_1^{\ \dot {\un{\a}}}\bar{\e}_2^{\ \dot 
{\un{\b}}}(R_{\dot {\un{\a}}\dot {\un{\b}}}^{\ \ \ \iota}+T_{\dot {\un{\a}}\dot {\un{\b}}}^{\ \ \un {c}}
\phi_{\un {c}}^{\ \iota}) \Big] ~~~.   \label{e3}
\end{align}

The supersymmetry variations of the inverse frame field $e_{\un  {a}}^{\ \un  {m}}(x)$,
gravitino $\psi_{\un  {a}}^{\ \un  {\d}}(x)$, and connection fields for the tangent space
symmetries $\phi_{\un  {a}}^{\ \iota} (x)$ take the forms below and are expressed in
terms dependent on linear and quadratic in the gravitino, and linear in the superspace 
torsions and curvature supertensors.  
\begin{align}
\d_{Q}e_{\un  {a}}^{\ \un  {m}} ~=&~ - \Big[~\e^{\un  {\b}}T_{\un  {\b}\un  {a}}^{\ \ \un  {d}} +\bar{\e
}^{\dot {\un{\b}}}T_{\dot {\un{\b}}\un  {a}}^{\ \ \un  {d}} +(\bar{\e}^{\dot {\un{\b}}}\psi_{\un  
{a}}^{\ \un  {\g}} + \e^{\un  {\g}}\bar{\psi}_{\un  {a}}^{\ \dot {\un{\b}}})T_{\dot {\un{\b}}\un  
{\g}}^{\ \ \un  {d}} \nonumber\\
& {~~~~~}
+ \e^{\un  {\b}}\psi_{\un  {a}}^{\ \un  {\g}}T_{\un  {\g}\un  {\b}}^{\ \ \un  
{d}} +\bar{\e}^{\dot {\un{\b}}}\bar{\psi}_{\un  {a}}^{\ \dot {\un{\g}}}T_{\dot {\un{\b}}\dot 
{\un{\g}}}^{\ \ \un  {d}}~\Big] \, e_{\un  {d}}^{\ \un  {m}} ~~~,
\label{e5} \\
\d_{Q}\psi_{\un  {a}}^{\ \un  {\d}} ~=&~ \textbf{{\rm D}}_{\un  {a}}\e^{\un  {\d}} - \e^{\un  {\b}}
(T_{\un  {\b}\un  {a}}^{\ \ \un  {\d}} + T_{\un  {\b}\un  {a}}^{\ \ \un  {e}}\psi_{\un  {e}}^{\ \un  
{\d}}) - \bar{\e}^{\dot {\un{\b}}}(T_{\dot {\un{\b}}\un  {a}}^{\ \ \un  {\d}} + T_{\dot {\un{\b}}\un  
{a}}^{\ \ \un  {e}}\psi_{\un  {e}}^{\ \un  {\d}}) \nonumber\\
& -(\bar{\e}^{\dot {\un{\b}}}\psi_{\un  {a}}^{\ \un  {\g}} + \e^{\un  {\g}}\bar{\psi}_{\un  {a}}^{\ 
\dot {\un{\b}}})(T_{\un  {\g}\dot {\un{\b}}}^{\ \ \un  {\d}} + T_{\un  {\g}\dot {\un{\b}}}^{\ \ \un  
{e}}\psi_{\un  {e}}^{\ \un  {\d}}) \nonumber\\
&  -\e^{\un  {\b}}\psi_{\un  {a}}^{\ \un  {\g}}(T_{\un  {\b}\un  {\g}}^{\ \ \un  {\d}} + T_{\un  {\b}
\un  {\g}}^{\ \ \un  {e}}\psi_{\un  {e}}^{\ \un  {\d}}) -\bar{\e}^{\dot {\un{\b}}}\bar{\psi}_{\un  
{a}}^{\ \dot {\un{\g}}}(T_{\dot {\un{\b}}\dot {\un{\g}}}^{\ \ \un  {\d}} + T_{\dot {\un{\b}}\dot 
{\un{\g}}}^{\ \ \un  {e}}\psi_{\un  {e}}^{\ \un  {\d}}) ~~~,   \label{e6} \\
\d_{Q}\phi_{\un  {a}}^{\ \iota} ~=&~ - \e^{\un  {\b}}(R_{\un  {\b}\un  {a}}^{\ \ \ \iota} + T_{\un  {\b}\un  
{a}}^{\ \ \un  {e}}\phi_{\un  {e}}^{\ \iota}) - \bar{\e}^{\dot {\un{\b}}}(R_{\dot {\un{\b}}\un  {a}}^{\ \ \ 
\iota} + T_{\dot {\un{\b}}\un  {a}}^{\ \ \un  {e}}\phi_{\un  {e}}^{\ \iota})  \nonumber\\
& -(\bar{\e}^{\dot {\un{\b}}}\psi_{\un  {a}}^{\ \un  {\g}} + \e^{\un  {\g}}\bar{\psi}_{\un  {a}}^{\ 
\dot {\un{\b}}})(R_{\un  {\g}\dot {\un{\b}}}^{\ \ \ \iota} + T_{\un  {\g}\dot {\un{\b}}}^{\ \ \un  {e}}
\phi_{\un  {e}}^{\ \iota}) \nonumber\\
&  -\e^{\un  {\b}}\psi_{\un  {a}}^{\ \un  {\g}}(R_{\un  {\b}\un  {\g}}^{\ \ \ \iota} + T_{\un  {\b}\un  
{\g}}^{\ \ \un  {e}}\phi_{\un  {e}}^{\ \iota}) -\bar{\e}^{\dot {\un{\b}}}\bar{\psi}_{\un  {a}}^{\ \dot 
{\un{\g}}}(R_{\dot {\un{\b}}\dot {\un{\g}}}^{\ \ \ \iota} + T_{\dot {\un{\b}}\dot {\un{\g}}}^{\ \ \un  
{e}}\phi_{\un  {e}}^{\ \iota}) ~~~.   \label{e7}
\end{align}

The supersymmetry covariantized versions of the torsions, gravitino field strength and field
strengths associated respective with the inverse frame field $e_{\un  {a}}^{\ \un  {m}}(x)$,
gravitino $\psi_{\un  {a}}^{\ \un  {\d}}(x)$, and connection fields for the tangent space
symmetries $\phi_{\un  {a}}^{\ \iota} (x)$ take the forms below and are expressed in
terms dependent on linear and quadratic in the gravitino, and linear in the superspace 
torsions and curvature supertensors.
\begin{align}
T_{\un  {a}\un  {b}}^{\ \ {\un c}} ~= &~ t_{\un  {a}\un  {b}}^{\ \ {\un c}} + \psi_{[\un  {a}}^{\ \un  {\d}} 
T_{\un  {\d}\un  {b}]}^{\ \ {\un c}} + \bar{\psi}_{[\un  {a}}^{\ \dot {\un{\d}}}T_{\dot {\un{\d}}\un  
{b}]}^{\ \ {\un c}} + \psi_{[\un  {a}}^{\ \un  {\d}}\bar{\psi}_{\un  {b}]}^{\ \dot {\un{\e}}}T_{\un  
{\d}\dot {\un{\e}}}^{\ \ {\un c}} + \psi_{\un  {a}}^{\ \un  {\d}}\psi_{\un  {b}}^{\ \un  {\e}}T_{\un  
{\d}\un  {\e}}^{\ \ {\un c}} +\bar{\psi}_{\un  {a}}^{\ \dot {\un{\d}}}\bar{\psi}_{\un  {b}}^{\ \dot 
{\un{\e}}} T_{\dot {\un{\d}}\dot {\un{\e}}}^{\ \ {\un c}} ~~~,  \label{e9}     \\
T_{\un  {a}\un  {b}}^{\ \ \g} ~= &~ t_{\un  {a}\un  {b}}^{\ \ \g} + \psi_{[\un  {a}}^{\ \un  {\d}} 
T_{\un  {\d}\un  {b}]}^{\ \ \g} + \bar{\psi}_{[\un  {a}}^{\ \dot {\un{\d}}}T_{\dot {\un{\d}}\un  
{b}]}^{\ \ \g} + \psi_{[\un  {a}}^{\ \un  {\d}}\bar{\psi}_{\un  {b}]}^{\ \dot {\un{\e}}}T_{\un  
{\d}\dot {\un{\e}}}^{\ \ \g} + \psi_{\un  {a}}^{\ \un  {\d}}\psi_{\un  {b}}^{\ \un  {\e}}T_{\un  
{\d}\un  {\e}}^{\ \ \g} +\bar{\psi}_{\un  {a}}^{\ \dot {\un{\d}}}\bar{\psi}_{\un  {b}}^{\ \dot 
{\un{\e}}} T_{\dot {\un{\d}}\dot {\un{\e}}}^{\ \ \g}  ~~~,  \label{e8}  \\
R_{\un  {a}\un  {b}}^{\ \ \iota} ~= &~ r_{\un  {a}\un  {b}}^{\ \ \iota} + \psi_{[\un  {a}}^{\ \un  
{\d}} R_{\un  {\d}\un  {b}]}^{\ \ \iota} + \bar{\psi}_{[\un  {a}}^{\ \dot {\un{\d}}}R_{\dot {\un{
\d}}\un  {b}]}^{\ \ \iota} + \psi_{[\un  {a}}^{\ \un  {\d}}\bar{\psi}_{\un  {b}]}^{\ \dot {\un{\e}}}
R_{\un  {\d}\dot {\un{\e}}}^{\ \ \ \iota} + \psi_{\un  {a}}^{\ \un  {\d}}\psi_{\un  {b}}^{\ \un  
{\e}}R_{\un  {\d}\un  {\e}}^{\ \ \iota} +\bar{\psi}_{\un  {a}}^{\ \dot {\un{\d}}}\bar{\psi}_{\un  
{b}}^{\ \dot {\un{\e}}} R_{\dot {\un{\d}}\dot {\un{\e}}}^{\ \ \iota} ~~~ .   \label{e10}
\end{align} 

In the linearized limit of these theories, not all of the terms in (\ref{e2}) - (\ref{e10}) appear.  
Instead these equations take the forms
\begin{align}
\xi^{\un  {m}} ~=&~ -\Big[(\e_1^{\ \un {\a}}\bar{\e}_2^{\ \dot {\un \b}}+\bar{\e}_1^{\ \dot {\un{\b}}}\e_2^{\ 
\un {\a}})T_{\un {\a} \dot {\un{\b}}}^{\ \ \un {c}} +\e_1^{\ \un  {\a}}\e_2^{\ \un  {\b}}T_{\un  {\a}\un 
{\b}}^{\ \ \un  {c}} + \bar{\e}_1^{\ \dot {\un{\a}}}\bar{\e}_2^{\ \dot {\un{\b}}}T_{\dot {\un{\a}}
\dot {\un{\b}}}^{\ \ \un  {c}}\Big]e_{\un  {c}}^{\ \un  {m}} ~~~,  \label{eF2} \\
\e^{\un  {\d}} ~=&~ -\Big[ (\e_1^{\ \un  {\a}}\bar{\e}_2^{\ \dot {\un{\b}}}+\bar{\e}_1^{\ \dot {\un{\b}}}
\e_2^{\ \un  {\a}})(T_{\un  {\a}\dot {\un{\b}}}^{\ \ \un  {\d}}+T_{\un \alpha\dot{\un \beta}}^{\ \ \un c}\psi_{\un c}^{\ \un \d}) + \e_1^{\ \un  {\a}}\e_2^{\ \un  {\b}}(
T_{\un  {\a}\un  {\b}}^{\ \ \un {\d}}+T_{\un  {\a}\un  {\b}}^{\ \ \un c}\psi_{\un c}^{\ \un \d}) + \bar{\e}_1^{\ \dot {\un{\a}}}\bar{\e}_2^{\ \dot {\un {\b}}}(T_{\dot {\un{\a}}\dot {\un{\b}}}^{\ \ \un  {\d}}+T_{\underline{\dot\a}\underline{\dot\b}}^{\ \ \un c}\psi_{\un c}^{\ \un \d})\Big] ~~~,  \label{eF4}  \\
\l^{\iota} ~=&~ -\Big[(\e_1^{\ \un  {\a}}\bar{\e}_2^{\ \dot {\un{\b}}}+\bar{\e}_1^{\ \dot {\un{\b}}}\e_2^{\ 
\un  {\a}})(R_{\un  {\a}\dot {\un{\b}}}^{\ \ \ \iota} +T_{\un\a\dot{\un\b}}^{\ \ \un c}\phi_{\un c}^{\ \iota}) + \e_1^{\ \un  {\a}}\e_2^{\ \un  {\b}}(R_{\un {\a}
\un  {\b}}^{\ \ \ \iota}+T_{\un\a\un\b}^{\ \ \un c}\phi_{\un c}^{\ \iota})  + \bar{\e}_1^{\ \dot {\un{\a}}}\bar{\e}_2^{\ \dot {\un{\b}}}(R_{\dot {\un{\a}}
\dot {\un{\b}}}^{\ \ \ \iota}+T_{\dot{\un\a}\dot{\un\b}}^{\ \ \un c}\phi_{\un c}^{\ \iota})\Big] ~~~,   \label{eF3} \\
\d_{Q}e_{\un  {a}}^{\ \un  {m}} ~=&~ -\Big[~\e^{\un  {\b}}T_{\un  {\b}\un  {a}}^{\ \ \un  {d}} +\bar{\e
}^{\dot {\un{\b}}}T_{\dot {\un{\b}}\un  {a}}^{\ \ \un  {d}} +(\bar{\e}^{\dot {\un{\b}}}\psi_{\un  
{a}}^{\ \un  {\g}} + \e^{\un  {\g}}\bar{\psi}_{\un  {a}}^{\ \dot {\un{\b}}})T_{\dot {\un{\b}}\un  
{\g}}^{\ \ \un  {d}} + \e^{\un  {\b}}\psi_{\un  {a}}^{\ \un  {\g}}T_{\un  {\g}\un  {\b}}^{\ \ \un  
{d}} +\bar{\e}^{\dot {\un{\b}}}\bar{\psi}_{\un  {a}}^{\ \dot {\un{\g}}}T_{\dot {\un{\b}}\dot 
{\un{\g}}}^{\ \ \un  {d}}\Big] \, e_{\un  {d}}^{\ \un  {m}} ~~~,
\label{eF5} \\
\d_{Q}\psi_{\un  {a}}^{\ \un  {\d}} ~=&~ \textbf{{\rm D}}_{\un  {a}}\e^{\un  {\d}} - \e^{\un  {\b}}
T_{\un  {\b}\un  {a}}^{\ \ \un  {\d}}  - \bar{\e}^{\dot {\un{\b}}}T_{\dot {\un{\b}}\un  {a}}^{\ \ \un  
{\d}}     ~~~,   
\label{eF6} \\
\d_{Q}\phi_{\un  {a}}^{\ \iota} ~=&~ - \e^{\un  {\b}}R_{\un  {\b}\un  {a}}^{\ \ \ \iota}  - \bar{\e}^{\dot 
{\un{\b}}}R_{\dot {\un{\b}}\un  {a}}^{\ \ \ 
\iota}    ~~~,   \label{eF7}  \\
T_{\un  {a}\un  {b}}^{\ \ \g} ~= &~ t_{\un  {a}\un  {b}}^{\ \ \g}  ~~~,  \label{eF8}  \\
T_{\un  {a}\un  {b}}^{\ \ {\un c}} ~= &~ t_{\un  {a}\un  {b}}^{\ \ {\un c}}  ~~~,  \label{eF9}     \\
R_{\un  {a}\un  {b}}^{\ \ \iota} ~= &~ r_{\un  {a}\un  {b}}^{\ \ \iota} ~~~ .   \label{eF10}
\end{align}
The terms on the RHS of the final three equation correspond to the non-supercovariantized 
versions of the respective torsions, gravitino field strength and connection field strengths.

\newpage
\section{Higher Dimensional Component Considerations}

In the following four subsections, we will appropriately adapt these results to the
cases of eleven and ten dimensional formulations appropriate for Nordstr\" om
supergravity in those contexts.  There are four steps:
\newline \indent
(a.) define a Nordstr\" om SG linearized superspace supercovariant derivative
in terms 
\newline \indent $~~~~~$
of a scalar prepotential leading to component fields,
\newline \indent
(b.) express the geometrical tensors of each respective superspace in 
terms of the 
\newline \indent $~~~~~$
component field presented in the previous part, 
\newline \indent
(c.) express the `composition rules' of the parameters of 
general coordinate,
local
\newline \indent $~~~~~$
Lorentz, and local SUSY transformations, and
\newline \indent
(c.) write the component level SUSY transformation laws
\vskip02.pt \noindent
that we undertake in each of the four cases of 11D, $\cal N$ = 1,
10D, $\cal N$ = 1, 10D, $\cal N$ = 2A, and 10D, $\cal N$ = 2B,
theories.

\subsection{Adaptation To 11D, $\cal N$ = 1 Component/Superspace Results: Step 1}

In the case of the 11D N(ordstr\" om)-SG covariant derivatives we define
\begin{align}
\nabla_{\a} ~=~& {\rm D}_{\a} + \frac{1}{2} \Psi {\rm D}_{\a} + \frac{1}{10} 
(\g^{\un{d}\un{e}})_{\a}{}^{\b} ({\rm D}_{\b}\Psi) {\cal M}_{\un{d}\un{e}}  ~~~, \\ 
\nabla_{\un{a}} ~=~& \pa_{\un{a}} + \Psi\pa_{\un{a}} + i \frac{1}{4} (\g_{\un{a}}
)^{\a\b} ({\rm D}_{\a}\Psi) {\rm D}_{\b} - (\pa_{\un{c}}\Psi) {\cal M}_{\un
{a}}{}^{\un{c}} ~~~,
\label{11d-1}
\end{align}
and ``split'' the spatial 11D N-SG covariant derivative into two parts
\begin{equation}
\nabla_{\un{a}}| ~=~ {\bf{D}}_{\un{a}} + \psi_{\un{a}}{}^{\g}
\nabla_{\g}|  ~~~.
\label{11d-2}
\end{equation}
On taking the $\theta$ $\to$ 0 limit the latter terms allows an identification with the gravitino
and the leading term in this limit yields a component-level linearized gravitationally covariant 
derivative operator given by
\begin{equation}
\begin{split}
{\bf{D}}_{\un{a}} ~=~& e_{\un{a}} + \phi_{\un a}{}^{\i} \mathcal{M}_{\i} 
~=~  \pa_{\un{a}} +\Psi \pa_{\un{a}} + \phi_{\un a}{}^{\i} \mathcal{M}_{\i} ~~~.
\end{split}
\label{11d-3}
\end{equation}
By comparison of the LHS to the RHS of (\ref{11d-3}), we see that a linearized frame field 
$e_{\un{a}} {}^{\un m}$ = $( \, 1 \,+\, \Psi)\d{}_{\un{a}} {}^{\un m}$ emerges to describe a 
scalar graviton.  Finally, comparison of the coefficient of the Lorentz generator $ \mathcal{
M}_{\i} $ as it appears in the latter two forms of (\ref{11d-3}) informs us the spin connection 
is given by
\begin{equation}
\phi_{\un{c}}{}^{\un{d}\un{e}} ~=~ - \frac{1}{2} \d_{\un{c}}{}^{[\un{d}} (\pa^{\un{e}]} 
\Psi) ~~~. 
\end{equation}
Comparing the result in (\ref{11d-1}) with the one in (\ref{11d-2}) a component gravitino 
is identified via
\begin{equation}
\psi_{\un{a}}{}^{\g} ~=~ i \frac{1}{4} (\g_{\un{a}})^{\g\d} ({\rm D}_{\d}\Psi) ~~~.
\end{equation}
However, as this expression contains an explicit $\g$-matrix we see that it really defines 
the non-conformal $\emph{spin-\fracm{1}{2}}$ part of the gravitino to be
\begin{equation}
    \psi_{\b} ~\equiv~ (\g^{\un{a}})_{\b\g} \psi_{\un{a}}{}^{\g}  ~~~.    
\end{equation}
This is to be expected. As a Nordstr\" om type theory only contains a scalar graviton,
it follows only the ``$\g$-trace'' of the gravitino can occur.  So then we have
\begin{equation}
{\rm D}_{\b}\Psi ~=~ i \frac{4}{11} (\g^{\un{a}})_{\b\g} \psi_{\un{a}}{}^{\g} ~\equiv~ 
i \frac{4}{11} \psi_{\b}  ~~~,
\end{equation}
in the $\theta$ $\to$ 0 limit.

In order to complete the specification of the geometrical superfields also requires explicit 
definitions of the bosonic terms to second order in D-derivatives.  So we define bosonic 
fields:
\begin{align}
K ~=~ C^{\g\d} ({\rm D}_{\g} {\rm D}_{\d} \Psi)  ~~~,~~~
K_{[3]} ~=~ (\g_{[3]})^{\g\d} ({\rm D}_{\g} {\rm D}_{\d} \Psi) ~~~,~~~
K_{[4]} ~=~ (\g_{[4]})^{\g\d} ({\rm D}_{\g} {\rm D}_{\d} \Psi)  ~~~,
\label{Fr1}
\end{align}
or in other words,
\begin{equation}
\frac{1}{2} {\rm D}_{[\g} {\rm D}_{\d]} \Psi ~=~ \frac{1}{32} \Big\{ C_{\g\d} K - 
\frac{1}{3!} (\g^{[3]})_{\g\d} K_{[3]} + \frac{1}{4!} (\g^{[4]})_{\g\d} K_{[4]} \Big\}
~~~. 
\label{Fr2}
\end{equation}
We emphasize that the component fields (the $K$'s) are defined by the $\theta$ 
$\to$ 0 limit of these equations.  The results in (\ref{Fr1}) and (\ref{Fr2}) follow 
as results from a Fierz identity
\be
\d{}_{[\g }{}^{\a} \d{}_{\d] }{}^{\b}
~=~  \frac{1}{16} \Big\{ C_{\g\d} C^{\a\b}  - 
\frac{1}{3!} (\g^{[3]})_{\g\d}  (\g_{[3]})^{\a\b} + \frac{1}{4!} (\g^{[4]})_{\g\d}  (\g_{[4]})^{\a\b}  
\Big\}
~~~,
\ee
valid for 11D spinors.
\subsection{Adaptation To 11D, $\cal N$ = 1 Component/Superspace Results: Step 2}

Torsions:
\begin{align}
T_{\a\b}^{\ \ \un{c}}  ~=~& i(\g^{\un{c}})_{\a\b} ~~~,  &&\\
T_{\a\b}^{\ \ \g} ~=~& i\frac{3}{110}(\g^{[2]})_{\a\b}(\g_{
[2]})^{\g\d}\psi_{\d}  ~~~, &&\\
T_{\a\un{b}}^{\ \ \un{c}} ~=~&  i\frac{3}{11} \Big[ \d_{\un b}^{\ \un c} \d_{\a}^{\ \b} 
+ \frac{3}{5} (\g_{\un b}^{\ \un c})_{\a}^{\ \b}  \Big] \psi_{\b}  ~~~,  &&\\
T_{\a\un{b}}^{\ \ \g} ~=~& i\frac{1}{128} \Big[ - (\g_{\un{b}})_{\a}^{\ 
\g} K + \frac{1}{2} (\g^{[2]})_{\a}^{\ \g} K_{\un{b}[2]} - \frac{1}{3!} 
(\g_{\un{b}[3]})_{\a}^{\ \g} K^{[3]}  + \frac{1}{3!} (\g^{[3]})_{\a
}^{\ \g} K_{\un{b}[3]} \nonumber\\
& \qquad  {~\,}  - \frac{1}{4!} (\g_{\un{b}[4]})_{\a}^{\ \g} K^{[4]}  \Big]  + \frac{1
}{8} \Big[ \d_{\un b}^{\ \un c} \d_{\a}^{\ \g} + 3 (\g_{\un b}^{\ \un c})_{\a}^{\ 
\g}  \Big] (\pa_{\un{c}}\Psi) ~~~,  &&\\
T_{\un{a}\un{b}}^{\ \ \un{c}} ~=~& 0 ~~~,  &&\\
T_{\un{a}\un{b}}^{\ \ \g} ~=~& \frac{1}{11}(\g_{[\un a})^{\g\d}(\pa_{
\un b]}\psi_{\d})  ~~~.
\end{align}

Curvatures:
\begin{align}
R_{\a\b}^{\ \ \ \un{d}\un{e}} ~=~& \frac{1}{80}  \Big[  (\g^{\un{d}\un{e}})_{\a
\b} K + (\g_{[1]})_{\a\b} K^{[1]\un{d}\un{e}}  - \frac{1}{3!} (\g^{\un{d}
\un{e}[3]})_{\a\b} K_{[3]} - \frac{1}{2} (\g_{[2]})_{\a\b} K^{[2]\un{d}\un{e
}}  \nonumber \\
& \qquad   + \frac{1}{5!4!} \e^{\un{d}\un{e}[5][4]} (\g_{[5]})_{\a\b} K_{[4]}   
\Big]  ~~~,  &&\\
R_{\a\un{b}}^{\ \ \ \un{d}\un{e}} ~=~&  i\frac{4}{11} \Big[ \d_{\un b}^{\ [\un d} (\pa^{
\un e]} \psi_{\a}) + \frac{1}{5} (\g^{\un{d}\un{e}})_{\a}^{\ \d}(\pa_{\un b}\psi_{\d}) 
\Big]  ~~~, &&\\
R_{\un{a}\un{b}}^{\ \ \ \un{d}\un{e}} ~=~& -(\pa_{[\un{a}}\pa^{[\un{d}}\Psi)\d_{
\un{b}]}^{\ \un{e}]} ~~~.
\end{align}

\subsection{Adaptation To 11D, $\cal N$ = 1 Component/Superspace Results: Step 3}

Parameter Composition Rules:
\begin{align}
\begin{split}
\xi^{\un{m}} ~=~&  - i \e_{1}{}^{\a} \e_{2}{}^{\b} (\g^{\un{c}})_{\a\b} \d_{\un{c}}{}^{\un 
{m}} (1+\Psi)  ~~~,
\end{split} \\
\begin{split}
\l^{\un{d}\un{e}} ~=~&  -  \frac{1}{80} \e_{1}{}^{\a} \e_{2}{}^{\b} \Big[  (\g^{\un{d}\un{
e}})_{\a\b} K + (\g_{[1]})_{\a\b} K^{[1]\un{d}\un{e}}  - \frac{1}{3!} (\g^{\un{d}\un{e}[3]}
)_{\a\b} K_{[3]} - \frac{1}{2} (\g_{[2]})_{\a\b} K^{[2]\un{d}\un{e}}  \\
& \qquad {~~~~~~~~~~}  + \frac{1}{5!4!} \e^{\un{d}\un{e}[5][4]} (\g_{[5]})_{\a\b} K_{[4]}   \Big]  
+ i \frac{1}{2} \e_{1}{}^{\a} \e_{2}{}^{\b} (\g^{[\un{d}})_{\a\b} (\pa^{\un{e}]} \Psi)  ~~~,
\end{split} \\
\begin{split}
\e^{\d} ~=~& i \frac{1}{11} \e_1^{\a}\e_2^{\b} \left[ (\g^{[1]})_{\a\b}(\g_{[1]
})^{\d\e} - \frac{3}{10} (\g^{[2]})_{\a\b}(\g_{[2]})^{\d\e} \right] \psi_{\e}  ~~~.
\end{split}
\end{align}

\subsection{Adaptation To 11D, $\cal N$ = 1 Component/Superspace Results: Step 4}

SUSY transformation laws:
\begin{align}
\begin{split}
\d_{Q} e_{\un{a}}{}^{\un{m}} ~=~& -i \frac{4}{11} \e^{\b} \left[ \d_{\un a}{}^{\un d
} \d_{\b}{}^{\g} + \frac{1}{5} (\g_{\un a}{}^{\un d})_{\b}{}^{ \g} \right] \d_{\un d}{}^{\un m} 
\psi_{\g}  ~~~,
\end{split} \\
\begin{split}
\d_{Q}\psi_{\un  {a}}{}^{\d} ~=~& (1 + \Psi) \pa_{\un{a}} \e^{\d} - \e^{\d} (\pa_{\un{c}} \Psi) 
\mathcal{M}_{\un{a}}{}^{\un{c}}  \\
& - i\frac{1}{128} \e^{\b} \Big[ - (\g_{\un{a}})_{\b}^{\ \d} K + \frac{1}{2} (\g^{[2]})_{\b}^{\ \d} 
K_{\un{a}[2]} - \frac{1}{3!} (\g_{\un{a}[3]})_{\b}^{\ \d} K^{[3]}  + \frac{1}{3!} (\g^{[3]})_{\b}^{
\ \d} K_{\un{a}[3]} \\
& \qquad  {~~~~~~~~} - \frac{1}{4!} (\g_{\un{a}[4]})_{\b}^{\ \d} K^{[4]}  \Big]  - \frac{1}{8} \e^{\b} \Big[ 
\d_{\un{a}}{}^{\un{c}} \d_{\b}^{\ \d} + 3 (\g_{\un{a}}{}^{\un{c}})_{\b}^{\ \d} \Big] (\pa_{\un{
c}}\Psi)  ~~~,
\end{split} \\
\begin{split}
\d_{Q} \phi_{\un  {a}}{}^{\un{d}\un{e}} ~=~&  -i \frac{4}{11} \e^{\b} \left[  \d_{\un a}
{}^{ [\un d} (\pa^{\un e]} \psi_{\b}) + \frac{1}{5}(\g^{\un{d}\un{e}})_{\b}{}^{ \d} (\pa_{\un 
a}\psi_{\d}) \right]   ~~~.
\end{split}
\end{align}

In the remaining subsections of the chapter, the steps described for the case of the 
11D, $\cal N$ = 1 theory above will be repeated, essentially line by line, in each of 
the cases for 10D, $\cal N$ = 1, 10D, $\cal N$ = 2A, and 10D, $\cal N$ = 2B superspaces.
This will imply a certain repetitive nature to the respective presentation.  There will
only be slight various in explicit details.  We are able to minimize this very slightly
by noting the result in (\ref{11d-3}) applies universally in all three cases.  So we will
not explicitly rewrite it nor its resultant implications several more times.

\subsection{Adaptation To 10D, $\cal N$ = 1 Component/Superspace Results: Step 1}

In the case of 10D $\mathcal{N} = 1$ N-SG covariant derivatives we define
\begin{align}  {~~~~~~~~~~}
\nabla_{\a}  ~=~& {\rm D}_{\a}+\frac{1}{2}\Psi {\rm D}_{\a}+\frac{1}{10}(\s^{\un{a}\un{
b}})_{\a}^{\ \b}({\rm D}_{\b}\Psi){\cal M}_{\un{a}\un{b}}  ~~~,  &&\\
\nabla_{\un{a}}  ~=~& \pa_{\un{a}} + \Psi\pa_{\un{a}} -i\frac{2}{5}(\s_{\un{a}})^{\a\b}
({\rm D}_{\a}\Psi){\rm D}_{\b} - (\pa_{\un{c}}\Psi){\cal M}_{\un{a}}{}^{\un{c}}  ~~~,
\label{10d1-1}
\end{align}
and ``split'' the spatial 10D $\mathcal{N} = 1$ N-SG covariant derivative into two parts
\begin{equation}
\nabla_{\un{a}}| ~=~ {\bf{D}}_{\un{a}} + \psi_{\un{a}}{}^{\g}
\nabla_{\g}|  ~~~.
\label{10d1-2}
\end{equation}
Comparing the result (\ref{10d1-1}) in with the one in (\ref{10d1-2}) a component gravitino 
is identified via
\begin{equation}
\psi_{\un{a}}{}^{\g} ~=~ - i \frac{2}{5} (\s_{\un{a}})^{\g\d} ({\rm D}_{\d}\Psi)  ~~~.
\end{equation}
However, as this expression contains an explicit $\s$-matrix we see that it defines 
the non-conformal $\emph{spin-\fracm{1}{2}}$ part of the gravitino to be
\begin{equation}
\psi_{\b} ~\equiv~ (\s^{\un{a}})_{\b\g} \psi_{\un{a}}{}^{\g} ~~~.
\end{equation}
and it follows only the ``$\s$-trace'' of the gravitino can occur.  So then we have
\begin{equation}
{\rm D}_{\b}\Psi ~=~ i \frac{1}{4} (\s^{\un{a}})_{\b\g} \psi_{\un{a}}{}^{\g} ~\equiv~ i \frac{1
}{4} \psi_{\b}  ~~~,
\end{equation}
in the $\theta$ $\to$ 0 limit.

The complete specification of the geometrical superfields also requires explicit 
definitions of the bosonic terms to second order in D-derivatives.  We take
advantage of the 10D Fierz identity
\be
\d{}_{[\g }{}^{\a} \d{}_{\d] }{}^{\b} ~=~  \frac{1}{48} \, (\s^{[3]})_{\g\d}  (\g_{[3]})^{\a\b} 
~~~,
\ee
valid for 10D spinors, so we may define a bosonic 
field:
\begin{align}
    G_{[3]} ~=~ (\s_{[3]})^{\g\d} ({\rm D}_{\g} {\rm D}_{\d} \Psi)  ~~~,
\end{align}
or in other words,
\begin{equation}
    \frac{1}{2} {\rm D}_{[\g} {\rm D}_{\d]} \Psi ~=~ \frac{1}{16\times 3!} (\s^{[3]})_{\g\d} G_{[3]} ~~~.
\end{equation}
We emphasize that the component field (the $G$) is defined by the $\theta$ 
$\to$ 0 limit of these equations.

\subsection{Adaptation To 10D, $\cal N$ = 1 Component/Superspace Results: Step 2}

Torsions:
\begin{align} 
T_{\a\b}^{\ \ \un{c}} ~=~ & i(\s^{\un{c}})_{\a\b} ~~~, &&\\
T_{\a\b}^{\ \ \g} ~=~ & 0 ~~~, &&\\
T_{\a\un{b}}^{\ \ \un{c}} ~=~ &  i\frac{3}{20} \left[
\d_{\un{b}}^{\ \un{c}}\d_{\a}^{\ \d} + (\s_{\un{b}}^{\ \un{c}})_{\a}^{\ \d
}\right] \psi_{\d}  ~~~, &&\\
T_{\a\un{b}}^{\ \ \g} ~=~ & i\frac{1}{80} \Big[-(\s^{[2]})_{\a}^{\ 
\g} G_{\un{b}[2]} + \frac{1}{3} (\s_{\un{b}[3]})_{\a}^{\ \g} G^{[3]} \Big]  - \frac{3}{10} \Big[ 
\d_{\un{b}}^{\ \un{c}} \d_{\a}^{\ \g} - (\s_{\un{b}}^{\ \un{c}})_{\a}^{\ \g} \Big] (\pa_{\un{c
}}\Psi)   ~~~, &&\\
T_{\un{a}\un{b}}^{\ \ \un{c}} ~=~ & 0 ~~~,  &&\\
T_{\un{a}\un{b}}^{\ \ \g} ~=~ & -\frac{1}{10}(\s_{[\un a})^{\g\d}(\pa_{\un b]}\psi_{\d}) ~~~.
\end{align}

Curvatures:
\begin{align} 
R_{\a\b}^{\ \ \ \un{d}\un{e}} ~=~ & -i\frac{6}{5}(\s^{[\un{d}})_{\a\b}
(\pa^{\un{e}]}\Psi) - \frac{1}{40} \Big[ \frac{1}{3!} (\s^{\un{d}\un{e}[3]})_{\a\b} G_{[3]} + 
(\s_{[1]})_{\a\b} G^{[1]\un{d}\un{e}} \Big]  ~~~,  &&\\
R_{\a\un{b}}^{\ \ \ \un{d}\un{e}} ~=~ & i \frac{1}{4} \Big[ \d_{\un b}^{\ [\un d}(\pa^{\un e]}
\psi_{\a}) +  \frac{1}{5}(\s^{\un{d}\un{e}})_{\a}^{\ \g}  (\pa_{\un b}\psi_{\g}) \Big]  
~~~, &&\\
R_{\un{a}\un{b}}^{\ \ \ \un{d}\un{e}} ~=~ & -(\pa_{[\un{a}}\pa^{[\un{d}}
\Psi)\d_{\un{b}]}^{\ \un{e}]}  ~~~.
\end{align}

\subsection{Adaptation To 10D, $\cal N$ = 1 Component/Superspace Results: Step 3}

Parameter Composition Rules:
\begin{align}
\xi^{\un{m}} ~=~& - i \e_{1}{}^{\a} \e_{2}{}^{\b} (\s^{\un{c}})_{\a\b} \d_{\un{c}}{}^{\un {m}} 
(1+\Psi)  ~~~, \\
\l^{\un{d}\un{e}} ~=~&   \frac{1}{40} \e_{1}{}^{\a} \e_{2}{}^{\b} \Big[  \frac{1}{3!} (\s^{
\un{d}\un{e}[3]})_{\a\b} G_{[3]} + (\s_{[1]})_{\a\b} G^{[1]\un{d}\un{e}}   \Big]  + i \frac{
17}{10} \e_{1}{}^{\a} \e_{2}{}^{\b} (\s^{[\un{d}})_{\a\b} (\pa^{\un{e}]} \Psi) ~~~, \\
\e^{\d} ~=~&   - i \frac{1}{10} \e_{1}{}^{\a} \e_{2}{}^{\b} (\s^{\un{c}})_{\a\b} (\s_{\un{c}})^{\d\e} 
\psi_{\e}   ~~~.
\end{align}
\subsection{Adaptation To 10D, $\cal N$ = 1 Component/Superspace Results: Step 4}

SUSY transformation laws:
\begin{align}
\d_{Q} e_{\un{a}}{}^{\un{m}} ~=~&  -i \frac{1}{4} \e^{\b} \left[ \d_{\un a}{}^{\un d} 
\d_{\b}{}^{\g} + \frac{1}{5} (\s_{\un a}{}^{ \un d})_{\b}{}^{\g}  \right] \d_{\un d}{}^{\un 
m} \psi_{\g} ~~~, \\
\begin{split}
\d_{Q}\psi_{\un  {a}}{}^{\d} ~=~& (1 + \Psi) \pa_{\un{a}} \e^{\d} - \e^{\d} (\pa_{\un{c}} \Psi) 
\mathcal{M}_{\un{a}}{}^{\un{c}}  \\
& - i\frac{1}{80} \e^{\b} \Big[-(\s^{[2]})_{\b}^{\ \d} G_{\un{a}[2]} + \frac{1}{3} (\s_{
\un{a}[3]})_{\b}^{\ \d} G^{[3]} \Big] + \frac{3}{10} \e^{\b} \Big[ \d_{\un{a}}^{\ \un{c}} \d_{\b}^{
\ \d} - (\s_{\un{a}}^{\ \un{c}})_{\b}^{\ \d} \Big] (\pa_{\un{c}}\Psi)  ~~~,
\end{split}  \\
\d_{Q} \phi_{\un  {a}}{}^{\un{d}\un{e}} ~=~& - i \frac{1}{4} \e^{\b} \left[  \d_{\un a
}^{\ [\un d} (\pa^{\un e]} \psi_{\b}) + \frac{1}{5}(\s^{\un{de}})_{\b}^{\ \g}(
\pa_{\un a}\psi_{\g}) \right]  ~~~.
\end{align}

\subsection{Adaptation To 10D, $\cal N$ = 2A Component/Superspace Results: Step 1}

In the case of 10D $\mathcal{N} = 2A$ N-SG covariant derivatives we define
\begin{align}  
\nabla_{\a} ~=~& {\rm D}_{\a} + \frac{1}{2}\Psi {\rm D}_{\a}+\frac{1}{10}(\s^{\un{
a}\un{b}})_{\a}^{\ \b}({\rm D}_{\b}\Psi){\cal M}_{\un{a}\un{b}}  ~~~, \\
\nabla_{\dot{\a}} ~=~& {\rm D}_{\dot{\a}} + \frac{1}{2}\Psi {\rm D}_{\dot{\a}}
+\frac{1}{10}(\s ^{\un{a}\un{b}})_{\dot{\a}}^{\ \dot{\b}}({\rm D}_{\dot{\b}}\Psi){\cal 
M}_{\un{a}\un{b}}  ~~~, \\
\nabla_{\un{a}} ~=~& \pa_{\un{a}}+\Psi\pa_{\un{a}} - i\frac{1}{5}(\s_{\un{a}})^{\d 
\g}({\rm D}_{\d}\Psi){\rm D}_{\g} - i\frac{1}{5}(\s_{\un{a}})^{\dot{\d} \dot{ \g}} 
({\rm D}_{\dot{\d}}\Psi){\rm D}_{\dot{\g}}-(\pa_{\un{c}}
\Psi){\cal M}_{\un{a}}^{\ \un{c}}  ~~~,
\label{10d2a-1}
\end{align}
and ``split'' the spatial 10D $\mathcal{N} = 2A$ N-SG covariant derivative into three parts
\begin{equation}
\nabla_{\un{a}}| ~=~ {\bf{D}}_{\un{a}} + \psi_{\un{a}}{}^{\g}
\nabla_{\g}| + \psi_{\un{a}}{}^{\dot{\g}} \nabla_{\dot{\g}}|  ~~~.
\label{10d2a-2}
\end{equation}
On taking the $\theta$ $\to$ 0 limit the latter terms allow an identification with
the component 
gravitinos are identified via
\begin{align}
\psi_{\un{a}}{}^{\g}  ~=~& - i \frac{1}{5} (\s_{\un{a}})^{\g\d} ({\rm D}_{\d}\Psi)  ~~~, ~~~
\psi_{\un{a}}{}^{\dot{\g}}  ~=~ - i \frac{1}{5} (\s_{\un{a}})^{\dot{\g}\dot{\d}} ({\rm D
}_{\dot{\d}}\Psi)  ~~~.
\end{align}
However, as this expression contains an explicit $\s$-matrix we see that it really defines 
the non-conformal $\emph{spin-\fracm{1}{2}}$ part of the gravitino to be
\begin{align}
\psi_{\b} ~\equiv~& (\s^{\un{a}})_{\b\g} \psi_{\un{a}}{}^{\g}  ~~~, ~~~
\psi_{\dot{\b}} ~\equiv~ (\s^{\un{a}})_{\dot{\b}\dot{\g}} \psi_{\un{a}}{}^{\dot{\g}}  ~~~.
\end{align}
It 
follows only the ``$\s$-trace'' of the gravitino can occur.  So then we have
\begin{align}
{\rm D}_{\b}\Psi ~=~ & i \frac{1}{2} (\s^{\un{a}})_{\b\g} \psi_{\un{a}}{}^{\g} ~\equiv~ i 
\frac{1}{2} \psi_{\b} ~~~, ~~~
{\rm D}_{\dot{\b}}\Psi ~=~  i \frac{1}{2} (\s^{\un{a}})_{\dot{\b}\dot{\g}} \psi_{\un{a}}
{}^{\dot{\g}} ~\equiv~ i \frac{1}{2} \psi_{\dot{\b}}  ~~~,
\end{align}
in the $\theta$ $\to$ 0 limit.

In order to complete the specification of the geometrical superfields also requires explicit 
definitions of the bosonic terms to second order in D-derivatives.  So we define bosonic 
fields:
\begin{align}
G_{[3]} ~=~& (\s_{[3]})^{\g\d} ({\rm D}_{\g} {\rm D}_{\d} \Psi) ~~~, ~~~
H_{[3]} ~=~ (\s_{[3]})^{\dot{\g}\dot{\d}} ({\rm D}_{\dot{\g}} {\rm D}_{\dot{\d}} \Psi)  ~~~,
\end{align}
\begin{align}
N ~=~& C^{\g\dot{\d}} ({\rm D}_{\g} {\rm D}_{\dot{\d}} \Psi)  ~~~, ~~~
N_{[2]} ~=~ (\s_{[2]})^{\g\dot{\d}} ({\rm D}_{\g} {\rm D}_{\dot{\d}} \Psi)  ~~~, ~~~
N_{[4]} ~=~ (\s_{[4]})^{\g\dot{\d}} ({\rm D}_{\g} {\rm D}_{\dot{\d}} \Psi)  ~~~,
\end{align}
or in other words,
\begin{align}
\frac{1}{2} {\rm D}_{[\g} {\rm D}_{\d]} \Psi ~=~& \frac{1}{16\times 3!} (\s^{[3]})_{\g\d} 
G_{[3]}   ~~~, ~~~
\frac{1}{2} {\rm D}_{[\dot{\g}} {\rm D}_{\dot{\d}]} \Psi ~=~ \frac{1}{16\times 3!} (\s^{
[3]})_{\dot{\g}\dot{\d}} H_{[3]}  ~~~,
\end{align}
and
\begin{equation}
{\rm D}_{\g} {\rm D}_{\dot{\d}} \Psi ~=~ \frac{1}{16} \Big\{ C_{\g\dot{\d}} N + \frac{1}{2!} 
(\s^{[2]})_{\g\dot{\d}} N_{[2]} + \frac{1}{4!} (\s^{[4]})_{\g\dot{\d}} N_{[4]} \Big\}  ~~~.
\end{equation}
We emphasize that the component fields (the $G$'s, $H$'s and $N$'s) are defined by 
the $\theta$ $\to$ 0 limit of these equations.

\subsection{Adaptation To 10D, $\cal N$ = 2A Component/Superspace Results: Step 2}

Torsions:
\begin{align}
T_{\a\b}^{\ \ \un{c}} ~=~ & i(\s^{\un{c}})_{\a\b} ~~~, &&\\
T_{\a\b}^{\ \ \g} ~=~ & i \frac{1}{10}(\s^{\un a})_{\a\b}(\s_{\un a})^{\g\d}\psi_{\d} 
~~~, &&\\
T_{\a\b}^{\ \ \dot\g} ~=~ & - i \frac{1}{10}(\s^{\un a})_{\a\b}(\s_{\un a})^{\dot\g
\dot\d}\psi_{\dot\d} ~~~, &&\\
T_{\dot\a\dot\b}^{\ \ \un{c}} ~=~ & i(\s^{\un{c}})_{\dot\a\dot\b} ~~~, &&\\
T_{\dot\a\dot\b}^{\ \ \g} ~=~ & -i\frac{1}{10}(\s^{\un a})_{\dot\a\dot\b}(\s_{\un a}
)^{\g\d}\psi_{\d} ~~~, &&\\
T_{\dot\a\dot\b}^{\ \ \dot\g} ~=~ & i\frac{1}{10}(\s^{\un a})_{\dot\a\dot\b}(\s_{\un 
a})^{\dot\g\dot\d}\psi_{\dot\d} ~~~, &&\\
T_{\a\dot\b}^{\ \ \un c} ~=~ & 0 ~~~, &&\\
T_{\a\dot\b}^{\ \ \g} ~=~ & i\frac{1}{4}\left[ \d_{\a}^{\ \g} \d_{\dot\b}^{\ \dot\d} + \frac{1}{10} 
(\s^{\un a \un b})_{\a}^{\ \g}(\s_{\un a\un b})_{\dot\b}^{\ \dot\d}\right] \psi_{\dot\d} 
~~~, &&\\
T_{\a\dot\b}^{\ \ \dot\g} ~=~ & i\frac{1}{4}\left[ \d_{\dot\b}^{\ \dot\g} \d_{\a}^{\ \d} + \frac{1}{
10} (\s^{\un a\un b})_{\dot\b}^{\ \dot\g} (\s_{\un a \un b})_{\a}^{\ \d} \right] 
\psi_{\d} ~~~, &&\\
T_{\a\un{b}}^{\ \ \un{c}} ~=~ & i\frac{1}{5}  \Big[ 2 \d_{\un{b}}^{\ \un{c}} \d_{\a}^{\ \d} + 
(\s_{\un{b}}^{\ \un{c}})_{\a}^{\ \d} \Big] \psi_{\d}  ~~~, &&\\
T_{\a\un{b}}^{\ \ \g} ~=~ & i\frac{1}{80} \Big[ -\frac{1}{2}(\s^{[2]})_{\a}^{\ 
\g} G_{\un{b}[2]} + \frac{1}{3!} (\s_{\un{b}[3]})_{\a}^{\ \g} G^{
[3]} \Big] - \frac{2}{5} \Big[ \d_{\un{b}}^{\ \un{c}} \d_{\a}^{\ \g}  - (\s_{\un{b}}^{\ \un{c}}
)_{\a}^{\ \g} \Big] (\pa_{\un{c}}\Psi)  ~~~, &&\\
T_{\a\un{b}}^{\ \ \dot\g} ~=~ & -i\frac{1}{80} \Big[(\s_{\un{b}})_{\a}^{\ 
\dot\g} N - (\s^{[1]})_{\a}^{\ \dot\g} N_{\un b[1]} + \frac{1}{2} (\s_{\un b[2]})_{\a}^{\ \dot\g} 
N^{[2]} - \frac{1}{3!}(\s^{[3]})_{\a}^{\ \dot\g} N_{\un b[3]}  \nonumber&&\\
& \qquad  {~~~~}  + \frac{1}{4!} (\s_{\un b[4]})_{\a}^{\ \dot\g} N^{[4]} \Big]    ~~~, &&\\
T_{\dot\a\un{b}}^{\ \ \un{c}} ~=~ & i \frac{1}{5} \Big[ 2 \d_{\un{b}}^{\ \un{c}} \d_{\dot\a}^{\ 
\dot\d} +  (\s_{\un{b}}^{\ \un{c}})_{\dot\a}^{\ \dot\d} \Big] \psi_{\dot\d} ~~~, &&\\
T_{\dot\a\un{b}}^{\ \ \g} ~=~ & -i\frac{1}{80} \Big[ (\s_{\un{b}})_{\dot\a}^{\ \g} N + (\s^{[1]}
)_{\dot\a}^{\ \g} N_{\un b[1]} - \frac{1}{2} (\s_{\un b[2]})_{\dot\a}^{\ \g} N^{[2]} - \frac{1}{3!} 
(\s^{[3]})_{\dot\a}^{\ \g} N_{\un b[3]}  \nonumber&&\\
& \qquad {~~~~}  + \frac{1}{4!} (\s_{\un b[4]})_{\dot\a}^{\ \g} N^{[4]} \Big]   ~~~, &&\\
T_{\dot\a\un{b}}^{\ \ \dot\g} ~=~ & i\frac{1}{80} \Big[ -\frac{1}{2} (\s^{[2]})_{\dot\a}^{\ 
\dot\g} H_{\un{b}[2]} + \frac{1}{3!} (\s_{\un{b}[3]})_{\dot\a}^{\ \dot\g} H^{[3]}  \Big]  - 
\frac{2}{5} \Big[ \d_{\un{b}}^{\ \un{c}} \d_{\dot\a}^{\ \dot\g} - (\s_{\un{b}}^{\ \un{c}}
)_{\dot\a}^{\ \dot\g} \Big] (\pa_{\un{c}}\Psi)   ~~~, &&\\
T_{\un{a}\un{b}}^{\ \ \un{c}} ~=~ & 0 ~~~,  &&\\
T_{\un{a}\un{b}}^{\ \ \g} ~=~ & -\frac{1}{10}(\s_{[\un{a
}})^{\g\d}(\pa_{\un{b}]}\psi_{\d}) ~~~, &&\\
T_{\un{a}\un{b}}^{\ \ \dot\g} ~=~ & -\frac{1}{10}(\s_{[\un{a
}})^{\dot\g\dot\d}(\pa_{\un{b}]}\psi_{\dot\d}) ~~~.
\end{align}

Curvatures:
\begin{align} 
R_{\a\b}^{\ \ \ \un{d}\un{e}} ~=~ & -i\frac{6}{5}(\s^{[\un{d}})_{\a\b} (\pa^{\un{e}]}\Psi) - 
\frac{1}{40} \left[\frac{1}{3!}(\s^{\un{d}\un{e}[3]})_{\a\b} G_{[3]} + (\s_{[1]})_{\a\b} 
G^{[1]\un{d}\un{e}} \right]  ~~~,  &&\\
R_{\dot\a\dot\b}^{\ \ \ \un{d}\un{e}} ~=~ & -i\frac{6}{5}(\s^{[\un{d}})_{\dot\a\dot\b}
(\pa^{\un{e}]}\Psi) - \frac{1}{40} \left[\frac{1}{3!} (\s^{\un{d}\un{e}[3]})_{\dot\a\dot\b} 
H_{[3]} + (\s_{[1]})_{\dot\a\dot\b} H^{[1] \un{d}\un{e}} \right]  ~~~,  &&\\
R_{\a\dot\b}^{\ \ \ \un{d}\un{e}} ~=~ & \frac{1}{40} \Big[ (\s^{\un{d}\un{e}})_{\a\dot{
\b}} N - C_{\a\dot{\b}} N^{\un{d}\un{e}} + \frac{1}{2} (\s^{\un{d}\un{e}[2]})_{\a\dot{
\b}} N_{[2]}   \nonumber\\
& \qquad  - \frac{1}{2} (\s_{[2]})_{\a\dot{\b}} N^{\un{d}\un{e}[2]}  + \frac{1}{4!4!} 
\e^{\un{d}\un{e}[4][\Bar{4}]} (\s_{[4]})_{\a\dot{\b}} N_{[\Bar{4}]}   \Big]    ~~~, &&\\
R_{\a\un{b}}^{\ \ \ \un{d}\un{e}} ~=~ & i\frac{1}{2} \Big[ \d_{\un{b}}^{\ [\un{d}}(\pa^{
\un{e}]}\psi_{\a}) + \frac{1}{5} (\s^{\un{d}\un{e}})_{\a
}^{\ \g}(\pa_{\un{b}}\psi_{\g}) \Big]  ~~~,  &&\\
R_{\dot\a\un{b}}^{\ \ \ \un{d}\un{e}} ~=~ & i\frac{1}{2} \Big[ \d_{\un{b}}^{\ [\un{d}}(
\pa^{\un{e}]}\psi_{\dot\a}) + \frac{1}{5}(\s^{\un{d}\un{e}})_{\dot\a
}^{\ \dot\g}(\pa_{\un{b}}\psi_{\dot\g}) \Big]  ~~~,  &&\\
R_{\un{a}\un{b}}^{\ \ \ \un{d}\un{e}} ~=~ & -(\pa_{[\un{a}}\pa^{[\un{d}}
\Psi)\d_{\un{b}]}^{\ \un{e}]}  ~~~.
\end{align}

\subsection{Adaptation To 10D, $\cal N$ = 2A Component/Superspace Results: Step 3}

Parameter Composition Rules:
\begin{align}
\xi^{\un{m}} ~=~& - i \big[~  \e_{1}{}^{\a} \e_{2}{}^{\b} (\s^{\un{c}})_{\a\b} + \e_{1}{}^{
\dot{\a}} \e_{2}{}^{\dot{\b}} (\s^{\un{c}})_{\dot{\a} \dot{\b}}  ~\big]~  \d_{\un{c}}{}^{\un 
{m}} (1 + \Psi) ~~~, \\
\begin{split}
\l^{\un{d}\un{e}} ~=~&  - \frac{1}{40} ( \e_{1}{}^{\a} \e_{2}{}^{\dot{\b}} + \e_{1}{}^{\dot{\b}} 
\e_{2}{}^{\a} ) \Big[ (\s^{\un{d}\un{e}})_{\a\dot{\b}} N - C_{\a\dot{\b}} N^{\un{d}\un{e}} + 
\frac{1}{2} (\s^{\un{d}\un{e}[2]})_{\a\dot{\b}} N_{[2]}  \\
& \qquad {~~~~~~~~~~~~~~~~~~~~~~~~} - \frac{1}{2} (\s_{[2]})_{\a\dot{\b}} N^{\un{
d}\un{e}[2]}  +  \frac{1}{4!4!} \e^{\un{d}\un{e}[4][\Bar{4}]} (\s_{[4]})_{\a\dot{\b}} 
N_{[\Bar{4}]}   \Big]  \\
& + \e_{1}{}^{\a} \e_{2}{}^{\b} \left[ i \frac{17}{10} (\s^{[\un{d}})_{\a\b} (\pa^{\un{
e}]}\Psi) + \frac{1}{40} \Big[ \frac{1}{3!}(\s^{\un{d}\un{e}[3]})_{\a\b} G_{[3]} + 
(\s_{[1]})_{\a \b} G^{[1]\un{d}\un{e}} \Big] \right]  \\
& + \e_{1}{}^{\dot{\a}} \e_{2}{}^{\dot{\b}} \left[ i \frac{17}{10} (\s^{[\un{d}})_{\dot{
\a}\dot{\b}} (\pa^{\un{e}]}\Psi) + \frac{1}{40} \Big[ \frac{1}{3!}(\s^{\un{d}\un{e}[3]}
)_{\dot{\a}\dot{\b}} H_{[3]} + (\s_{[1]})_{\dot{\a}\dot{\b}} H^{[1]\un{d}\un{e}} \Big] 
\right]  ~~~,
\end{split}  \\
\begin{split}
\e^{\d} ~=~& - i\frac{1}{4}( \e_{1}{}^{\a} \e_{2}{}^{\dot{\b}} + \e_{1}{}^{\dot{\b}} \e_{2}{}^{
\a} ) \left[ \d_{\a}^{\ \d} \d_{\dot{\b}}^{\ \dot{\e}}  + \frac{1}{10} (\s^{[2]})_{\a}^{\ \d}(
\s_{[2]})_{\dot\b}^{\ \dot\e} \right] \psi_{\dot\e}   \\
& \qquad - i \frac{1}{5} \e_{1}{}^{\a} \e_{2}{}^{\b} (\s^{\un c})_{\a\b} (\s_{\un c})^{
\d\e}\psi_{\e}  ~~~.
\end{split}
\end{align}

\subsection{Adaptation To 10D, $\cal N$ = 2A Component/Superspace Results: Step 4}

SUSY transformation laws:
\begin{align}
\d_{Q} e_{\un{a}}{}^{\un{m}} ~=~&  -i \frac{1}{2} \e^{\b} \left[ \d_{\un a}{}^{\un d} 
\d_{\b}{}^{\g} + \frac{1}{5} (\s_{\un a}{}^{ \un d})_{\b}{}^{\g}  \right] \d_{\un d}{}^{\un 
m} \psi_{\g} -i \frac{1}{2} \e^{\dot{\b}} \left[ \d_{\un a}{}^{\un d} \d_{\dot{\b}}{}^{
\dot{\g}} + \frac{1}{5} (\s_{\un a}{}^{ \un d})_{\dot{\b}}{}^{\dot{\g}}  \right] \d_{\un 
d}{}^{\un m} \psi_{\dot{\g}}  ~~~, \\
\begin{split}
\d_{Q}\psi_{\un  {a}}{}^{\d} ~=~& (1 + \Psi) \pa_{\un{a}} \e^{\d} - \e^{\d} (\pa_{\un{c}} \Psi) 
\mathcal{M}_{\un{a}}{}^{\un{c}}  \\
& - i\frac{1}{80} \e^{\b} \Big[ - \frac{1}{2} (\s^{[2]})_{\b}{}^{\d} G_{\un{a}[2]} + \frac{1}{
3!} (\s_{\un{a}[3]})_{\b}{}^{\d} G^{[3]} \Big] + \frac{2}{5} \e^{\b} \Big[ \d_{\un{a}}{}^{\un{
c}} \d_{\b}{}^{\d} - (\s_{\un{a}}{}^{\un{c}})_{\b}{}^{\d} \Big] (\pa_{\un{c}}\Psi)    \\
& + i\frac{1}{80} \e^{\dot {\b}} \Big[ (\s_{\un{a}})_{\dot\b}{}^{\d} N + (\s^{[1]})_{\dot\b}{}^{
\d} N_{\un{a}[1]} - \frac{1}{2} (\s_{\un{a}[2]})_{\dot\b}{}^{\d} N^{[2]} - \frac{1}{3!} (\s^{[3]})_{
\dot\b}{}^{\d} N_{\un{a}[3]} \\
& \qquad {~~~~~~} + \frac{1}{4!} (\s_{\un{a}[4]})_{\dot\b}{}^{\d} N^{[4]} \Big]  ~~~,
\end{split}  \\
\d_{Q} \phi_{\un  {a}}{}^{\un{d}\un{e}} ~=~&  - i\frac{1}{2} \e^{\b} \Big[  \d_{\un{a}}^{\ [\un{d}}
(\pa^{\un{e}]}\psi_{\b}) + \frac{1}{5}(\s^{\un{d}\un{e}})_{\b}^{\ \g}(\pa_{\un{a}}\psi_{\g})  
\Big]   - i\frac{1}{2} \e^{\dot{\b}} \Big[  \d_{\un{a}}^{\ [\un{d}}(\pa^{\un{e}]}\psi_{\dot\b}) + 
\frac{1}{5}(\s^{\un{d}\un{e}})_{\dot\b}^{\ \dot\g}(\pa_{\un{a}}\psi_{\dot\g})  \Big] ~~~.
\end{align}

\subsection{Adaptation To 10D, $\cal N$ = 2B Component/Superspace Results: Step 1}

In the case of 10D $\mathcal{N} = 2B$ N-SG covariant derivatives we define
\begin{align}  
\nabla_{\a} ~=~& {\rm D}_{\a} + \frac{1}{2} \Psi {\rm D}_{\a} + \frac{1}{10} (\s^{\un{a}
\un{b}})_{\a}{}^{\b} ({\rm D}_{\b}\Psi) {\cal M}_{\un{a}\un{b}}  ~~~,  \\
\Bar{\nabla}_{\a} ~=~& \Bar{\rm D}_{\a} + \frac{1}{2}\Bar{\Psi} \Bar{\rm D}_{\a} + \frac{1}{10} 
(\s^{\un{a}\un{b}})_{\a}{}^{\b} (\Bar{\rm D}_{\b}\Bar{\Psi}) {\cal M}_{\un{a}\un{b}}  ~~~, \\
\begin{split}
\nabla_{\un{a}} ~=~& \pa_{\un{a}} + \frac{1}{2}\Psi\pa_{\un{a}} + \frac{1}{2}\Bar{\Psi} \pa_{
\un{a}} - i\frac{1}{32} (\s_{\un{a}})^{\a \b} ({\rm D}_{\a}\Bar{\Psi}) \Bar{\rm D}_{\b} - 
i\frac{1}{32} (\s_{\un{a}})^{\a \b} (\Bar{\rm D}_{\a} \Psi) {\rm D}_{\b} \\
& - i\frac{27}{160} (\s_{\un{a}})^{\a \b} ({\rm D}_{\a}\Psi) \Bar{\rm D}_{\b} - i\frac{27}{
160} (\s_{\un{a}})^{\a \b} (\Bar{\rm D}_{\a}\Bar{\Psi}) {\rm D}_{\b}  \\
& -\frac{1}{2}(\pa_{\un{c}}\Psi){\cal M}_{\un{a}}{}^{\un{c}} - \frac{1}{2} (\pa_{\un{c}}\Bar{
\Psi}) {\cal M}_{\un{a}}{}^{\un{c}} ~~~,
\end{split}
\label{10d2b-1}
\end{align}
and ``split'' the spatial 10D $\mathcal{N} = 2B$ N-SG covariant derivative into three parts
\begin{equation}
\nabla_{\un{a}}| ~=~ {\bf{D}}_{\un{a}} + \psi_{\un{a}}{}^{\g}
\nabla_{\g}| + \Bar{\psi}_{\un{a}}{}^{\g}\Bar{\nabla}_{\g}|  ~~~.
\label{10d2b-2}
\end{equation}
On taking the $\theta$ $\to$ 0 limit the latter terms allows an identification with the 
gravitino and the leading term in this limit yields a component-level linearized 
gravitationally covariant derivative operator given by
\begin{equation}
\begin{split}
{\bf{D}}_{\un{a}} ~=~ e_{\un{a}} + \phi_{\un a}{}^{\i} \mathcal{M}_{\i}  ~=~ 
\pa_{\un{a}} + \frac{1}{2}(\Psi+\Bar{\Psi}) \pa_{\un{a}} + \phi_{\un a}{}^{\i} 
\mathcal{M}_{\i}  ~~~.
\end{split}
\label{10d2b-3}
\end{equation}
Comparison of the LHS to the RHS of (\ref{10d2b-3}), we see that a linearized frame 
field $e_{\un{a}} {}^{\un m}$ = $( \, 1 \,+\, \frac{1}{2}(\Psi+\Bar{\Psi}) \,)\d{}_{\un{a}} {}^{
\un m}$ emerges to describe a scalar graviton.  Finally, comparison of the coefficient 
of the Lorentz generator $ \mathcal{M}_{\i} $ as it appears in the latter two forms of 
(\ref{10d2b-3}) informs us the spin connection is given by
\begin{equation}
\phi_{\un{c}}{}^{\un{d}\un{e}} ~=~ - \frac{1}{4} \d_{\un{c}}{}^{[\un{d}} \big( \pa^{\un{e}]} 
( \Psi + \Bar{\Psi} ) \big)  ~~~.
\end{equation}
Comparing the result (\ref{10d2b-1}) in with the one in (\ref{10d2b-2}) the component 
gravitinos are identified via
\begin{align}
\psi_{\un{a}}{}^{\g} ~=~& - i \frac{1}{160} (\s_{\un{a}})^{\g\d} \big( \Bar{\rm D}_{\d} 
( 5 \Psi + 27 \Bar{\Psi} )  \big)  ~~~, \\
\Bar{\psi}_{\un{a}}{}^{\g} ~=~& - i \frac{1}{160} (\s_{\un{a}})^{\g\d} \big( {\rm D}_{
\d} ( 5 \Bar{\Psi} + 27 \Psi ) \big)  ~~~,
\end{align}
which are equivalent to
\begin{align}
\Bar{\rm D}_{\a} ( 5 \Psi + 27 \Bar{\Psi} ) ~=~& i 16 (\s^{\un a})_{\a\g} \psi_{\un a}
{}^{\g}  ~~~,~~~
{\rm D}_{\a} ( 5 \Bar{\Psi} + 27 \Psi ) ~=~ i 16 (\s^{\un a})_{\a\g} \Bar{\psi}_{\un 
a}^{\ \g}  ~~~.  \label{10d2b-4b}
\end{align}
However, as this expression contains an explicit $\s$-matrix we see that it really defines 
the non-conformal $\emph{spin-\fracm{1}{2}}$ part of the gravitino to be
\begin{align}
\psi_{\b} ~\equiv~& (\s^{\un{a}})_{\b\g} \psi_{\un{a}}{}^{\g}  ~~~, ~~~
\Bar{\psi}_{\b} ~\equiv~ - (\s^{\un{a}})_{\b\g} \Bar{\psi}_{\un{a}}{}^{\g}  ~~~.
\end{align}
Since the results in 
(\ref{10d2b-4b}) are under-constrained, we are allowed to introduce a fermionic 
auxiliary field $\l_{\a}$ and its complex conjugate $\Bar{\l}_{\a}$. So 
then we have
\begin{align}
\Bar{\rm D}_{\a}\Psi ~=~& i \frac{1}{2} (\s^{\un a})_{\a\g} \psi_{\un a }{}^{\g} - 27 
\Bar{\l}_{\a} ~\equiv~ i \frac{1}{2} \psi_{\a} - 27 \Bar{\l}_{\a} ~~~, \\
\Bar{\rm D}_{\a} \Bar{\Psi} ~=~& i \frac{1}{2} (\s^{\un a})_{\a\g} \psi_{\un a }{}^{\g} 
+ 5 \Bar{\l}_{\a}  ~\equiv~ i \frac{1}{2} \psi_{\a} + 5 \Bar{\l}_{\a}  ~~~, \\
{\rm D}_{\a} \Bar{\Psi} ~=~& i \frac{1}{2} (\s^{\un a})_{\a\g} \Bar{\psi}_{\un a }{}^{\g} 
- 27 \l_{\a}  ~\equiv~ - i \frac{1}{2} \Bar{\psi}_{\a} - 27 \l_{\a}  ~~~, \\
{\rm D}_{\a}\Psi ~=~& i \frac{1}{2} (\s^{\un a})_{\a\g} \Bar{\psi}_{\un a }{}^{\g} 
+ 5 \l_{\a}  ~\equiv~ - i \frac{1}{2} \Bar{\psi}_{\a} + 5 \l_{\a} ~~~,\label{cnGt}
\end{align}
in the $\theta$ $\to$ 0 limit. Also observe that 
\begin{align}
\Bar{\rm D}_{\a} (\Bar{\Psi} - \Psi) ~=~& 32 \Bar{\l}_{\a}  ~~~, ~~~
{\rm D}_{\a} (\Psi - \Bar{\Psi}) ~=~ 32 \l_{\a}  ~~~.
\end{align}

In order to complete the specification of the geometrical superfields also requires 
explicit definitions of the bosonic terms to second order in D-derivatives.  So we 
define bosonic fields:
\begin{align}
U_{[3]} ~=~& (\s_{[3]})^{\g\d} ({\rm D}_{\g} {\rm D}_{\d} \Psi) ~~~, &   \Bar{U}_{[3]} 
~=~& - (\s_{[3]})^{\g\d} (\Bar{\rm D}_{\g} \Bar{\rm D}_{\d} \Bar{\Psi}) ~~~, \\
X_{[3]} ~=~& (\s_{[3]})^{\g\d} (\Bar{\rm D}_{\g} \Bar{\rm D}_{\d} \Psi) ~~~, & \Bar{
X}_{[3]} ~=~& - (\s_{[3]})^{\g\d} ({\rm D}_{\g} {\rm D}_{\d} \Bar{\Psi}) ~~~,  \\
Y_{[3]} ~=~& (\s_{[3]})^{\g\d} ({\rm D}_{\g} \Bar{\rm D}_{\d} \Psi)  &  \Bar{Y}_{[3]} 
~=~& - (\s_{[3]})^{\g\d} (\Bar{\rm D}_{\g} {\rm D}_{\d} \Bar{\Psi}) \nonumber\\
~=~& (\s_{[3]})^{\g\d} (\Bar{\rm D}_{\g} {\rm D}_{\d} \Psi) ~~~,  &  ~=~& - (\s_{[3]
})^{\g\d} ({\rm D}_{\g} \Bar{\rm D}_{\d} \Bar{\Psi})  ~~~.
\end{align}
Ue emphasize that the component fields (the $U$'s, $X$'s and $Y$'s) are defined 
by the $\theta$ $\to$ 0 limit of these equations.

\subsection{Adaptation To 10D, $\cal N$ = 2B Component/Superspace Results: Step 2}

Torsions:
\begin{align}
T_{\a\b}^{\ \ \un{c}} ~=~ & 0 ~~~, &&\\
T_{\a\b}^{\ \ \g} ~=~ & - i\frac{1}{5}(\s^{\un c})_{\a\b}(\s_{\un c})^{\g\d}\Bar{
\psi}_{\d} + 2(\s^{\un c})_{\a\b}(\s_{\un c})^{\g\d} \l_{\d} ~~~, &&\\
T_{\a\b}^{\ \ \bar\g} ~=~ & 0 ~~~, &&\\
T_{\bar\a\bar\b}^{\ \ \un{c}} ~=~ & 0 ~~~, &&\\
T_{\bar\a\bar\b}^{\ \ \g} ~=~ & 0 ~~~, &&\\
T_{\bar\a\bar\b}^{\ \ \bar\g} ~=~ & i\frac{1}{5}(\s^{\un c})_{\a\b}(\s_{\un c})^{
\g\d} \psi_{\d} + 2(\s^{\un c})_{\a\b}(\s_{\un c})^{\g\d} \Bar{\l}_{\d} ~~~, 
&&\\
T_{\a\bar\b}^{\ \ \un{c}} ~=~ & i(\s^{\un c})_{\a\b} ~~~, &&\\
T_{\a\bar\b}^{\ \ \g} ~=~ & -i \frac{1}{240}  (\s^{[3]})_{\a\b}(\s_{[3]})^{\g\d}
\psi_{\d} + \frac{1}{8} \Big[ (\s^{[3]})_{\a\b}(\s_{[3]})^{\g\d} - \frac{1}{30}(
\s^{[5]})_{\a\b}(\s_{[5]})^{\g\d} \Big] \Bar{\l}_{\d} ~~~, \\
T_{\a\bar\b}^{\ \ \bar\g} ~=~ & i \frac{1}{240}  (\s^{[3]})_{\a\b}(\s_{[3]})^{\g
\d} \Bar{\psi}_{\d} + \frac{1}{8} \Big[ (\s^{[3]})_{\a\b}(\s_{[3]})^{\g\d} - \frac{
1}{30}(\s^{[5]})_{\a\b}(\s_{[5]})^{\g\d} \Big] \l_{\d}  ~~~, \\
T_{\a\un b}^{\ \ \un c}  ~=~ & - i \frac{1}{5} \Big[ 2 \d_{\un b}^{\ \un c}\d_{\a}^{\ \g} + 
(\s_{\un b}^{\ \un c})_{\a}^{\ \g} \Big] \Bar{\psi}_{\g} + \Big[ - 11 \d_{\un b}^{\ 
\un c}\d_{\a}^{\ \g} + (\s_{\un b}^{\ \un c})_{\a}^{\ \g} \Big] \l_{\g}   ~~~,  
&&\\
T_{\a\un b}^{\ \ \g} ~=~ & \frac{1}{64} \Big[ -31 \d_{\un b}^{\ \un c} \d_{\a}^{\ \g} + 15 
(\s_{\un b}^{\ \un c})_{\a}^{\ \g} \Big] (\pa_{\un c}\Psi) + \frac{1}{320} \Big[ 27 
\d_{\un b}^{\ \un c} \d_{\a}^{\ \g}  + 53 (\s_{\un b}^{\ \un c})_{\a}^{\ \g} \Big] (
\pa_{\un c}\Bar\Psi) &&\nonumber\\ 
& -i\frac{1}{2560}\Big[\frac{1}{2}(\s^{[2]})_{\a}^{\ \g} \Big( 5 Y_{\un b[2]} - 27 
\Bar{Y}_{\un b[2]} \Big) - \frac{1}{3!} (\s_{\un b [3]})_{\a}^{\ \g} \Big( 5 Y^{[3]} - 
27 \Bar{Y}^{[3]} \Big) \Big]  ~~~, &&\\
T_{\a\un b}^{\ \ \bar\g} ~=~ & -i\frac{1}{2560}\Big[\frac{1}{2}(\s^{[2]})_{\a}^{\ \g} 
\Big( - 5 \Bar{X}_{\un b[2]} + 27 U_{\un b[2]} \Big) - \frac{1}{3!} (\s_{\un b [3]}
)_{\a}^{\ \g} \Big( - 5 \Bar{X}^{[3]} + 27 U^{[3]} \Big) \Big]  ~~~, &&\\
T_{\bar\a\un b}^{\ \ \un c}  ~=~ & i \frac{1}{5} \Big[ 2 \d_{\un b}^{\ \un c}\d_{\a}^{\ \g} 
+ (\s_{\un b}^{\ \un c})_{\a}^{\ \g} \Big] \psi_{\g} + \Big[ - 11 \d_{\un b}^{\ \un c}
\d_{\a}^{\ \g} + (\s_{\un b}^{\ \un c})_{\a}^{\ \g} \Big] \Bar{\l}_{\g} ~~~, &&\\
T_{\bar\a\un b}^{\ \ \g} ~=~ & -i\frac{1}{2560}\Big[\frac{1}{2}(\s^{[2]})_{\a}^{\ \g} 
\Big( 5 X_{\un b[2]} - 27 \Bar{U}_{\un b[2]} \Big) - \frac{1}{3!} (\s_{\un b [3]})_{
\a}^{\ \g} \Big( 5 X^{[3]} - 27 \Bar{U}^{[3]} \Big) \Big]  ~~~, &&\\
T_{\bar\a\un b}^{\ \ \bar\g} ~=~ \nonumber&  \frac{1}{64} \Big[ -31  \d_{\un b}^{\ \un 
c}\d_{\a}^{\ \g} + 15 (\s_{\un b}^{\ \un c})_{\a}^{\ \g} \Big] (\pa_{\un c}\Bar\Psi) 
+ \frac{1}{320} \Big[ 27 \d_{\un b}^{\ \un c} \d_{\a}^{\ \g} + 53 (\s_{\un b}^{\ \un 
c})_{\a}^{\ \g} \Big] (\pa_{\un c}\Psi)  &&\\
& -i\frac{1}{2560}\Big[\frac{1}{2}(\s^{[2]})_{\a}^{\ \g} \Big( - 5 \Bar{Y}_{\un b[2]} 
+ 27 Y_{\un b[2]} \Big) - \frac{1}{3!} (\s_{\un b [3]})_{\a}^{\ \g} \Big( - 5 \Bar{Y
}^{[3]} + 27 Y^{[3]} \Big) \Big]  ~~~, &&\\
T_{\un{a}\un{b}}^{\ \ \un c} ~=~ & 0 ~~~, &&\\
T_{\un{a}\un{b}}^{\ \ \g} ~=~ & -\frac{1}{10}(\s_{[\un{a
}})^{\g\d}(\pa_{\un{b}]}\psi_{\d})  ~~~, &&\\
T_{\un{a}\un{b}}^{\ \ \bar\g} ~=~ & \frac{1}{10}(\s_{[\un{a
}})^{\g\d}(\pa_{\un{b}]}\Bar{\psi}_{\d}) ~~~.
\end{align}

Curvatures:
\begin{align}
R_{\a\b}^{\ \ \ \un{d}\un{e}} ~=~ & \frac{1}{40}\left[\frac{1}{3!}(\s^{\un{d}\un{e}[3]}
)_{\a\b} U_{[3]} - (\s_{[1]})_{\a\b} U^{[1] \un{d}\un{e}} \right]   ~~~,  &&\\
R_{\bar\a\bar\b}^{\ \ \ \un{d}\un{e}} ~=~ & - \frac{1}{40}\left[\frac{1}{3!}(\s^{\un{d}
\un{e}[3]})_{\a\b} \Bar{U}_{[3]} - (\s_{[1]})_{\a\b} \Bar{U}^{[1] \un{d}\un{e}} \right]  
~~~,  &&\\
R_{\a\bar\b}^{\ \ \ \un{d}\un{e}} ~=~ \nonumber& -i\frac{3}{5}(\s^{[\un d})_{\a\b}
(\pa^{\un e]}(\Psi+\Bar\Psi)) - i\frac{1}{10}(\s^{\un{d}\un{e}\un{f}})_{\a\b}(\pa_{
\un f}(\Psi+\Bar\Psi)) &&\\
& -\frac{1}{80} \Big[ (\s_{[1]})_{\a\b} \Big( Y^{[1]\un{d}\un{e}} - \Bar{Y}^{[1]\un{d}
\un{e}} \Big) - \frac{1}{2} (\s^{[2][\un{d}})_{\a\b} \Big( Y^{\un{e}]}_{\ \ [2]} - \Bar{Y
}^{\un{e}]}_{\ \ [2]} \Big)   ~~~  &&\\
& {~~~~~~~~\,} - \frac{1}{3!} (\s^{\un{d}\un{e}[3]})_{\a\b} \Big( Y_{[3]} 
- \Bar{Y}_{[3]} \Big) \Big] ~~~,  &&\\
R_{\a\un{b}}^{\ \ \ \un{d}\un{e}} ~=~ & - i \frac{1}{2} \Big[ \d_{\un{b}}^{\ [\un{d}} (\pa^{
\un{e}]} \Bar{\psi}_{\a} ) + \frac{1}{5} (\s^{\un{d}\un{e}})_{\a}^{\ \g} (\pa_{\un{b}} \Bar
{\psi}_{\g}) \Big]  - 11  \d_{\un{b}}^{\ [\un{d}} (\pa^{\un{e}]} \l_{\a} ) +  (\s^{\un{
d}\un{e}})_{\a}^{\ \g} (\pa_{\un{b}} \l_{\g})   ~~~,  &&\\
R_{\bar\a\un{b}}^{\ \ \ \un{d}\un{e}} ~=~ & i \frac{1}{2} \Big[ \d_{\un{b}}^{\ [\un{d}} (
\pa^{\un{e}]} \psi_{\a} ) + \frac{1}{5} (\s^{\un{d}\un{e}})_{\a}^{\ \g} (\pa_{\un{b}} \psi_{
\g}) \Big]  - 11  \d_{\un{b}}^{\ [\un{d}} (\pa^{\un{e}]} \Bar{\l}_{\a} ) +  (\s^{\un{
d}\un{e}})_{\a}^{\ \g} (\pa_{\un{b}} \Bar{\l}_{\g}) ~~~,  &&\\
R_{\un{a}\un{b}}^{\ \ \ \un{d}\un{e}} ~=~ & - \frac{1}{2} \big( \pa_{[\un{a}}\pa^{[\un{
d}} (\Psi + 
\Bar\Psi) \big) \d_{\un{b}]}^{\ \un{e}]}   ~~~.
\end{align}

\subsection{Adaptation To 10D, $\cal N$ = 2B Component/Superspace Results: Step 3}

Parameter Composition Rules:
\begin{align}
\xi^{\un{m}} ~=~& - i ( \e_{1}{}^{\a} \Bar{\e}_{2}{}^{\b} + \Bar{\e}_{1}{}^{\b} \e_{2}{}^{\a} ) 
(\s^{\un{c}})_{\a\b} \d_{\un c}{}^{\un m} \Big( 1 + \frac{1}{2} (\Psi + \Bar{\Psi} ) \Big)  ~~~,  \\
\begin{split}
\l^{\un{d}\un{e}} ~=~& - ( \e_{1}{}^{\a} \Bar{\e}_{2}{}^{\b} + \Bar{\e}_{1}{}^{\b} \e_{2}{}^{\a} )  
\bigg[  - i \frac{17}{20}(\s^{[\un d})_{\a\b}(\pa^{\un e]}(\Psi+\Bar\Psi)) - i\frac{1}{10} 
(\s^{\un{d}\un{e}\un{f}})_{\a\b}(\pa_{\un f}(\Psi+\Bar\Psi))  \\
& \qquad {~~~~~~~~~~~~~~~~~~~~~} -\frac{1}{80} \Big[ (\s_{[1]})_{\a\b} \Big( 
Y^{[1]\un{d}\un{e}} - \Bar{Y}^{[1] \un{d}\un{e}} \Big) - \frac{1}{2} (\s^{[2][\un{d}})_{\a
\b} \Big( Y^{\un{e}]}_{\ \ [2]} - \Bar{Y}^{\un{e}]}_{\ \ [2]} \Big)     \\
& \qquad {~~~~~~~~~~~~~~~~~~~~~}  - \frac{1}{3!} (\s^{\un{d}\un{e}[3]})_{\a\b} \Big( 
Y_{[3]} - \Bar{Y}_{[3]} \Big) \Big] ~ \bigg]  \\
& - \frac{1}{40} \e_{1}{}^{\a} \e_{2}{}^{\b} \left[\frac{1}{3!}(\s^{\un{d}\un{e}[3]})_{\a\b} 
U_{[3]} - (\s_{[1]})_{\a\b} U^{[1] \un{d}\un{e}} \right]   \\
&+ \frac{1}{40} \Bar{\e}_{1}{}^{\a} 
\Bar{\e}_{2}{}^{\b} \left[\frac{1}{3!}(\s^{\un{d}\un{e}[3]})_{\a\b} \Bar{U}_{[3]} - 
(\s_{[1]})_{\a\b} \Bar{U}^{[1] \un{d}\un{e}} \right]  ~~~,
\end{split} \\
\begin{split}
\e^{\d} ~=~&  - ( \e_{1}{}^{\a} \Bar{\e}_{2}{}^{\b} + \Bar{\e}_{1}{}^{\b} \e_{2}{}^{\a} ) \bigg[  
i \frac{1}{10} \Big[ (\s^{[1]})_{\a\b}(\s_{[1]})^{\d\e} - \frac{1}{24} (\s^{[3]})_{
\a\b}(\s_{[3]})^{\d\e}  \Big] \psi_{\e} \\
& \qquad + \frac{1}{8} \Big[ (\s^{[3]})_{\a\b}(\s_{[3]})^{\d\e} - \frac{1}{30}(\s^{
[5]})_{\a\b}(\s_{[5]})^{\d\e} \Big] \Bar{\l}_{\e}  \bigg]  \\
& - \e_{1}{}^{\a} \e_{2}{}^{\b}  \left[  - i\frac{1}{5}(\s^{[1]})_{\a\b}(\s_{[1]})^{\d\e} 
\Bar{\psi}_{\e} + 2(\s^{[1]})_{\a\b}(\s_{[1]})^{\d\e} \l_{\e}   \right]  ~~~.
\end{split}
\end{align}

\subsection{Adaptation To 10D, $\cal N$ = 2B Component/Superspace Results: Step 4}

\begin{align}
\begin{split}
\d_{Q} e_{\un{a}}{}^{\un{m}} ~=~&  - \e^{\b} \left[ - i \frac{1}{2} \Big[ \d_{\un a}{}^{\un d} 
\d_{\b}{}^{\g} + \frac{1}{5} (\s_{\un a}{}^{ \un d})_{\b}{}^{\g} \Big] \Bar{\psi}_{\g} +  
\Big[ -11 \d_{\un a}{}^{\un d} \d_{\b}{}^{\g} +  (\s_{\un a}{}^{ \un d})_{\b}{}^{\g} \Big] 
\l_{\g} \right] \d_{\un d}{}^{\un m}  \\
& - \Bar{\e}^{\b} \left[ i \frac{1}{2} \Big[ \d_{\un a}{}^{\un d} \d_{\b}{}^{\g} + \frac{1}{5} 
(\s_{\un a}{}^{ \un d})_{\b}{}^{\g} \Big] \psi_{\g} +  \Big[ -11 \d_{\un a}{}^{\un d} 
\d_{\b}{}^{\g} +  (\s_{\un a}{}^{ \un d})_{\b}{}^{\g} \Big] \Bar{\l}_{\g} \right]
\d_{\un d}{}^{\un m}   ~~~,
\end{split} \\
\begin{split}
\d_{Q}\psi_{\un  {a}}{}^{\d} ~=~& \Big( 1 + \frac{1}{2} (\Psi + \Bar{\Psi} ) \Big) \pa_{
\un{a}} \e^{\d} - \frac{1}{2} \e^{\d} (\pa_{\un{c}} (\Psi+\Bar{\Psi})) \mathcal{M}_{\un{a}}{}^{
\un{c}}  \\
&  -  \frac{1}{64} \e^{\b} \Big[ -31 \d_{\un a}^{\ \un c} \d_{\b}^{\ \d} + 15 (\s_{\un a
}^{\ \un c})_{\b}^{\ \d} \Big] (\pa_{\un c}\Psi) -  \frac{1}{320} \e^{\b} \Big[ 27 \d_{\un a}^{
\ \un c} \d_{\b}^{\ \d}  + 53 (\s_{\un a}^{\ \un c})_{\b}^{\ \d} \Big] (\pa_{\un c}\Bar\Psi) \\
& + i \frac{1}{2560} \e^{\b} \Big[ \frac{1}{2}(\s^{[2]})_{\b}^{\ \d} \Big( 5 Y_{\un a[2]} 
- 27 \Bar{Y}_{\un a[2]} \Big) - \frac{1}{3!} (\s_{\un a [3]})_{\b}^{\ \d} \Big( 5 Y^{[3]} 
- 27 \Bar{Y}^{[3]} \Big) \Big]  \\
& + i \frac{1}{2560} \Bar{\e}^{\b} \Big[\frac{1}{2}(\s^{[2]})_{\b}^{\ \d} \Big( 5 X_{\un 
a[2]} - 27 \Bar{U}_{\un a[2]} \Big) - \frac{1}{3!} (\s_{\un a [3]})_{\b}^{\ \d} \Big( 5 
X^{[3]} - 27 \Bar{U}^{[3]} \Big) \Big]  ~~~,
\end{split}  \\
\begin{split}
\d_{Q} \phi_{\un  {a}}{}^{\un{d}\un{e}} ~=~&  i \frac{1}{2} \e^{\b} \Big[  \d_{\un{a}}^{\ [\un
{d}} (\pa^{\un{e}]} \Bar{\psi}_{\b} ) +  \frac{1}{5} (\s^{\un{d}\un{e}})_{\b}^{\ \g} (\pa_{\un{a}} 
\Bar{\psi}_{\g}) \Big] - \e^{\b} \Big[ - 11  \d_{\un{a}}^{\ [\un{d}} (\pa^{\un{e}]} \l_{\b} ) 
+  (\s^{\un{d}\un{e}})_{\b}^{\ \g} (\pa_{\un{a}} \l_{\g})  \Big]  \\
& - i \frac{1}{2} \Bar{\e}^{\b} \Big[ \d_{\un{a}}^{\ [\un{d}} (\pa^{\un{e}]} \psi_{\b} ) + \frac{1}
{5} (\s^{\un{d}\un{e}})_{\b}^{\ \g} (\pa_{\un{a}} \psi_{\g}) \Big] -   \Bar{\e}^{\b} \Big[ - 11  
\d_{\un{a}}^{\ [\un{d}} (\pa^{\un{e}]} \Bar{\l}_{\b} ) +  (\s^{\un{d}\un{e}})_{\b}^{\ 
\g} (\pa_{\un{a}} \Bar{\l}_{\g}) \Big]  ~~~.
\end{split}
\end{align}

\newpage
\section{10D, $\cal N$ = 2B Chiral Compensator Considerations}

In the limits where all supergravity fields are set to zero, four sets of super algebras emerge.
These take the forms: \newline \noindent
(a.) 11D, $\cal N$ = 1,
\be
 \left\{ \, {\rm D}_{\a} ~,~ {\rm D}_{\b} \, \right\} ~=~ i\, (\g{}^{\un a}){}_{\a \b}  \, \pa_{\un a}
 ~~,~~
  \left[ \, {\rm D}_{\a} ~,~ {\pa}_{\un b} \, \right] ~=~ 0 ~~,~~ 
   \left[ \, {\pa}_{\un a} ~,~ {\pa}_{\un b} \, \right] ~=~ 0
\ee
(b.) 10D, $\cal N$ = 1,
\be 
 \left\{ \, {\rm D}_{\a} ~,~ {\rm D}_{\b} \, \right\} ~=~ i\, (\s{}^{\un a}){}_{\a \b}  \, \pa_{\un a}
 ~~,~~
\left[ \, {\rm D}_{\a} ~,~ {\pa}_{\un b} \, \right] ~=~ 0 ~~,~~ 
\left[ \, {\pa}_{\un a} ~,~ {\pa}_{\un b} \, \right] ~=~ 0    
\ee
(c.) 10D, $\cal N$ = 2A,
\be  \eqalign{
{~~~~~~}
&\left\{ \, {\rm D}_{\a} ~,~ {\rm D}_{\b} \, \right\} ~=~ i\, (\s{}^{\un a}){}_{\a \b}  \, \pa_{\un a}
 ~~,~~
 \left\{ \, {\rm D}_{\dot \a} ~,~ {\rm D}_{\dot \b} \, \right\} ~=~ i\, (\s{}^{\un a}){}_{{\dot \a} {\dot \b}}  
 \, \pa_{\un a}  
  ~~,~~
 \left\{ \, {\rm D}_{ \a} ~,~ {\rm D}_{\dot \b} \, \right\} ~=~ 0   ~~, \cr
 &{~~\,}\left[ \, {\rm D}_{\a} ~,~ {\pa}_{\un b} \, \right] ~=~0
 ~~,~~{~~~~~~~~~~~~\,~~}
 \left[ \, {\rm D}_{\dot \a} ~,~ {\pa}_{\un b} \, \right] ~=~ 0 
  ~~,~~ {~~~~~~~~~~\,~~~\,~~}
 \left[ \, {\pa}_{ \un a} ~,~ {\pa}_{\un b} \, \right] ~=~ 0   ~~, 
   } \ee
(d.) 10D, $\cal N$ = 2B,
\be  \eqalign{
{~~~~~~}
&\left\{ \, {\rm D}_{\a} ~,~ {\rm D}_{\b} \, \right\} ~=~ 0  ~~,~~
 \left\{ \, \Bar{\rm D}_{\a} ~,~  \Bar{\rm D}_{\b} \, \right\} ~=~0    ~~,~~
 \left\{ \, {\rm D}_{ \a} ~,~  \Bar{\rm D}_{\b} \, \right\} ~=~  i\, (\s{}^{\un a}){}_{\a \b}  \, \pa_{\un a}   ~~, \cr
 &{~~\,}\left[ \, {\rm D}_{\a} ~,~ {\pa}_{\un b} \, \right] ~=~0
 ~~,~~{~~~}
 \left[ \,  \Bar{\rm D}_{\a} ~,~ {\pa}_{\un b} \, \right] ~=~ 0 
  ~~,~~ {~~~~}
 \left[ \, {\pa}_{ \un a} ~,~ {\pa}_{\un b} \, \right] ~=~ 0   ~~, 
   } \ee
   
We next introduce a complex superfield denoted by $\O{}_{d}$ into each of these $d$-dimensional
superspaces and seek to probe the implications of impose a first order differential equation imposed on
this superfield that utilizes any of the spinorial derivatives above.   

For either the 11D, $\cal N$ = 1 or 10D, $\cal N$ = 1 superspaces we have
\be 
{\rm D}_{ \b} \, \O{}_{d} ~=~ 0 ~~\to~~ {\rm D}_{ \a} {\rm D}_{ \b} \, \O{}_{d} ~=~ 0
~~\to~~ \{ \, {\rm D}_{ \a} ~,~ {\rm D}_{ \b} \} \,  \O{}_{d} ~=~ 0 ~~\to~~ \pa{}_{ \un c} \, \O{}_{d} ~=~ 0
~~~,
\label{Trv1}
\ee
and by analogy for the  10D, $\cal N$ = 2A superspace we find
$$
{\rm D}_{ \b} \, \O{}_{d} ~=~ 0 ~~\to~~ {\rm D}_{ \a} {\rm D}_{ \b} \, \O{}_{d} ~=~ 0
~~\to~~ \{ \, {\rm D}_{ \a} ~,~ {\rm D}_{ \b} \} \,  \O{}_{d} ~=~ 0 ~~\to~~ \pa{}_{ \un c} \, \O{}_{d} ~=~ 0
~~~,
$$
\be {~~~~~\,}
{\rm D}_{\dot \b} \, \O{}_{d} ~=~ 0 ~~\to~~ {\rm D}_{\dot \a} {\rm D}_{\dot \b} \, \O{}_{d} ~=~ 0
~~\to~~ \{ \, {\rm D}_{\dot \a} ~,~ {\rm D}_{\dot \b} \} \,  \O{}_{d} ~=~ 0 ~~\to~~ \pa{}_{ \un c} \, \O{}_{d} ~=~ 0
~~~,
\label{Trv2}
\ee
Thus, from (\ref{Trv1}) and (\ref{Trv2}) we find the superfield  $\O{}_{d}$ in each of these $d$-dimensional
superspaces must be a constant.  However, upon repeating these considerations for the  10D, $\cal N$ = 2B 
superspace we find
\be \eqalign{
{\rm D}_{ \b} \, \O{}_{d} ~&=~ 0 ~~\to~~ {\rm D}_{ \a} {\rm D}_{ \b} \, \O{}_{d} ~=~ 0
~~\to~~ \{ \, {\rm D}_{ \a} ~,~ {\rm D}_{ \b} \} \,  \O{}_{d} ~=~ 0 ~~\to~~ 0 ~=~ 0
~~~,  \cr
 \Bar{\rm D}_{\b} \, \O{}_{d} ~&=~ 0 ~~\to~~  \Bar{\rm D}_{\a}  \Bar{\rm D}_{\b} \, \O{}_{d} ~=~ 0
~~\to~~ \{ \,  \Bar{\rm D}_{\a} ~,~  \Bar{\rm D}_{\b} \} \,  \O{}_{d} ~=~ 0 ~~\to~~ 0 ~=~ 0
~~~,}
\label{Trv3}
\ee
which shows that the superfield $\O{}_{d}$ in this case can be a non-trivial representation of
the translation operator.  

The differential equation
\be
 \Bar{\rm D}_{\b} \, \O{}_{d} ~=~ 0  ~~~,
\ee
in the context of four dimensions implies that $\O{}_{d}$ is a ``chiral superfield.''  On the other
hand the differential equation
\be
{\rm D}_{\b} \, \O{}_{d} ~=~ 0 ~~~,
\ee
in the context of four dimensions implies that $\O{}_{d}$ is a ``anti-chiral superfield.''    While it
is not possible to simultaneously impose both conditions because a chiral superfield is the
complex conjugate of an anti-chiral one, either one or the other can be imposed.  This also
means that neither the chiral nor the anti-chiral condition can be applied to a real superfield.

Let us return to the results shown (\ref{cnGt}) by focusing only on the  equations that
contain $ \l_{\a} $
\be \eqalign{
{\rm D}_{\a}\Psi ~=~& i \frac{1}{2} (\s^{\un a})_{\a\g} \Bar{\psi}_{\un a }{}^{\g} 
+ 5 \l_{\a}  ~\equiv~ - i \frac{1}{2} \Bar{\psi}_{\a} + 5 \l_{\a} ~~~, \cr
{\rm D}_{\a} \Bar{\Psi} ~=~& i \frac{1}{2} (\s^{\un a})_{\a\g} \Bar{\psi}_{\un a }{}^{\g} 
- 27 \l_{\a}  ~\equiv~ - i \frac{1}{2} \Bar{\psi}_{\a} - 27 \l_{\a}   ~~~,} 
\label{cnGt2}
\ee
since the remaining equations can be obtain by complex conjugation.  In all the other
cases we have explored, there is no spinor field such as $ \l_{\a} $.  Taking the difference of
the two equations that appear in (\ref{cnGt2}), we may obtain
\be
i \, \fracm 1{32} \,
{\rm D}_{\a} {\big ( } \, \Psi ~-~  {\Bar \Psi}  \, {\big ) } ~=~ i\,  \l_{\a}   ~~~.
\ee
However, the quantity $i  \,  (  \Psi \,-\, {\Bar \Psi} ) $ is a real superfield.  The requirement
that $ \l_{\a} $ = 0 is equivalent to the imposition of an anti-chirality condition on a
real superfield and this condition possesses no non-trivial solution.

The inability to introduce such a chiral superfield distinguishes the type 2B theory from
the other higher dimensional constructions we have considered.  At first order in the
$\theta$-expansion of $\Psi$ both the spin-1/2 portion of the gravitino ${\Bar \psi}_{\un 
a }{}^{\g} $ {\em {and}} a separate spin-1/2 auxiliary spinor $ \l_{\a} $ must exist.

\newpage

\section{Conclusion \& Possible Future Directions}
\label{conclusions}

 \vskip,2in

In this work, we have presented the forms of the superspace torsions and curvature 
supertensors that are consistent with Nordstr\" om supergravity in eleven and ten 
dimensional superspaces.  For the superspaces in 11D, $\cal N$ = 1, 10D, $\cal N$ = 
1,  10D, $\cal N$ = 2A, and  10D, $\cal N$ = 2B, these results are found in the sets 
of equations given as (3.12) - (3.20), (3.36) - (3.44), (3.62) - (3.85), and (3.110) -
(3.134), respectively.  To our knowledge, these presentations initiate new results
for the superspace torsions and curvature supertensors in these domains.

The use of the superfield $\Psi$ in all cases guarantees all of these theories
are ``off-shell'' supersymmetric without the need to impose some equations
of motion for the fulfillment of a local supersymmetry algebra.  The fact that
$\Psi$ used in each case does not satisfy any a priori superdifferential constraint
implies the closure.  Unfortunately, this same fact also implies that each of the
descriptions we have provided is not an irreducible one.  Exploring the possibility
of imposing further superdifferential constraints to obtain one or more irreducible
representations is the work for the future.

The work completed in this paper also suggests two new pathways to explore 
elements of 11D, $\cal N$ = 1 supergeometry.

\noindent
(a.) \newline \indent
In the works of \cite{2MT,2MTcf} on the basis of the study of solutions to the 11D
superspace Bianchi identities up to engineering dimension one, forms for the 
superspace torsions and curvature supertensors were proposed.  Upon comparing 
particularly the results in the first of these references to the result derived in the 
current work as seen in (3.12) - (3.20), apparent concurrence is found.  In the work 
of \cite{2MT}, we can use the definition\footnote{In comparison to these older works
we have ``rescaled'' $t_{[2]} $, $U_{[3]}$, $V_{[4]} $, and $Z_{[5]}$ relative to the
original definitions. }
\be
\nabla_{\a} J_{\b} ~=~ C_{\a\b} S + (\g^{\un{a}})_{\a\b} v_{\un{a}} + \frac{1}{2} (\g^{
[2]})_{\a\b} t_{[2]} + \frac{1}{3!} (\g^{[3]})_{\a\b} U_{[3]} + \frac{1}{4!} (\g^{[4]})_{\a\b} 
V_{[4]} + \frac{1}{5!} (\g^{[5]})_{\a\b} Z_{[5]}  ~~~.  \label{delJdef}
\ee
In this former work, we must set the 11D ``on-shell'' superfield $W{}_{\un a \un b 
\un c \un d}$ to zero to make comparisons.  When this is done, then by a change of 
notation where
\be
\psi{}_{\a}  ~\to~ J {}_{\a} ~~~,~~~ K{}_{\un a}  ~\to~ v {}_{\un a} ~~~,~~~
K{}_{[2]}  ~\to~ t {}_{[2]} ~~~,~~~ K{}_{[3]}  ~\to~ U{}_{[3]} ~~~,~~~
K{}_{[4]}  ~\to~ V{}_{[4]} ~~~,~~~ K{}_{[5]}  ~\to~ Z{}_{[5]} ~~~,~~~
\ee
we then look at (\ref{delJdef}) in contrast to the form of (3.8) and
(3.9) in this work.  We find in the Nordstr\" om limit,
\be
v {}_{\un a} ~=~ \pa{}_{\un a} \Psi ~~~,~~~
t {}_{[2]} ~=~ 0 ~~~,~~~
 Z{}_{[5]} ~= ~ 0 ~~~,
 \label{fieldsolns}
\ee
and thus there is significant overlap.  In particular, the results in (\ref{fieldsolns}) tell 
us something interesting about the $J{}_{\a}$ tensor.  We can decompose it into two 
parts
\be
J{}_{\a} ~=~ J{}_{\a}^{(T)} ~+~ {\rm D}{}_{\a} \Psi
\ee
which is equivalent to the usual decomposition of a gauge field into its transverse
and longitudinal parts.  Upon setting the $J{}_{\a}^{(T)}$ = 0, one recovers the 
Nordstr\" om theory.

There is a further feature noted in the work of  \cite{2MTcf} that also is indicated as 
a direction to include in this new pathway of exploration for 11D superspace
supergravity.

While the notation of superconformal symmetry is not presently understood in a
number of approaches to the study of 11D supergravity, the superspace approach
in \cite{2MTcf} is indicative of a specific further modification.  In particular, by the
introduction of a scaling transformation of the supervielbein, it was found that
a modification of the spinor-spinor-vector component of the supertorsion that is
given by the expression
\be
T{}_{\a \b}{}^{\un c} ~=~ i(\g^{\un{c}})_{\a\b} 
~+~ i(\g^{[\un 2]})_{\a\b} \, {\cal X}{}_{[{\un 2}]}{}^{\un c}
~+~ i(\g^{[\un 5]})_{\a\b}  \, {\Hat {\cal X}}{}_{[{\un 5}]}{}^{\un c}
\ee
is consistent with the superspace scale transformations if and only if the ``$\cal 
X$-tensor'' and ``$\Hat {\cal X}$-tensor'' satisfy the conditions,
\be
{\cal X}{}_{{\un a}{\un c}}{}^{\un c} ~=~ 0  ~~~,~~~ \e{}^{[\un 8]  {\un a}{\un b} {\un c}}
{\cal X}{}_{{\un a}{\un b}}{}_{\un c} ~=~ 0 ~~~,~~~ 
{\Hat {\cal X}}{}_{[{\un 4}]{\un c}}{}^{\un c} ~=~ 0  ~~~,~~~ \e{}^{[\un 5]  {\un a}{\un b} {\un c}
{\un d}{\un e} {\un f}}
{\Hat {\cal X}}{}_{{\un a}{\un b}{\un c}{\un d}{\un e}}{}_{\un f} ~=~ 0  ~~~.
\ee
A detailed and careful study of the 11D superspace supergravity Bianchi identities 
with the modifications in the current work as well as the works of \cite{2MT,2MTcf} 
is indicated to assess the form of any equations of motion that emerges in the presence 
of retaining the on-shell field strength.

\noindent
(b.) \newline \indent
While the pathway for future investigation described above depends on the
study of 11D supergravity supercovariant tensors and their Bianchi identities,
the ``Breitenlohner Approach'' suggests a second pathway.  

The 4D, $\cal N$
= 1 Wess-Zumino gauge vector supermultiplet in (\ref{V1}) (or alternately the 
component level 4D, $\cal N$ = 1 supermultiplet) arises in a very interesting 
way related to the  4D, $\cal N$ = 1 real pseudoscalar superfield $V$.   The
components fields in $V$ may be expressed as an expansion in terms of
the fermionic superspace D-operators followed by taking the limit as $\q{}^a$
goes to zero.  See equation (4.3.4a) in \cite{SpRSp8BK} and the equivalent
expressions using the Majorana superspace coordinates associated with
the superspace relevant to the component results in (\ref{V1}) take the forms
\be  \eqalign{ {~~~~~~~~}
&C ~=~ V \big| ~~~,~~~ 
\chi{}_a ~=~ {\rm D}{}_a V \big| ~~~,~~~ 
M ~=~ C{}^{a b} {\rm D}{}_a  {\rm D}{}_b V \big| ~~~,~~~ 
N ~=~ i (\g{}^5){}^{a b} {\rm D}{}_a  {\rm D}{}_b V \big| ~~~,~~~     \cr
&v{}_{\un a} ~=~ (\g{}^5 \g{}_{\un a}){}^{a b} {\rm D}{}_a  {\rm D}{}_b V \big| ~~~,~~~ 
\l{}^a  ~=~ \e{}^{a \, b \,c \, d} {\rm D}{}_b  {\rm D}{}_c{\rm D}{}_d V \big| ~~~,~~~ 
{\rm d}  ~=~ \e{}^{a \, b \,c \, d} {\rm D}{}_a {\rm D}{}_b  {\rm D}{}_c{\rm D}{}_d V \big|
~~~, \label{fielddefs}
}\ee
where $ \e{}^{a \, b \,c \, d}$ is the Levi-Civita tensor defined over the
Majorana spinor indices.  Also we have made adaptations in the notation 
that are appropriate for Majorana basis conventions in 4D.  The results in 
(\ref{fielddefs}) make clear there are eight bosons and eight fermions contained in this 
superfield.  It is also clear there is a component level gauge 1-form $v{}_{
\un a}$ that occurs at the quadratic order in the $\q$-expansion of $V$.  

Now let us consider the situation of an 11D, $\cal N$ = 1 scalar superfield
${\cal V}{}^{(11)}$ analogous to $V$.  There are some differences of course.  For 
example, in ${\cal V}{}^{(11)}$ there are 2,147,483,648 bosonic component fields 
and 2,147,483,648 fermionic component fields.  In the 11D, $\cal N$ = 1 
superspace, the quadratic order spinor supercovariant derivatives are
in the $\{ 1\}$, \{165\}, and \{330\} representations of the 11D Lorentz
group whose explicit forms are
\be  \eqalign{
\D{}^{(1)} ~&=~ C^{\a \b} \, {\rm D}_{\a} \, {\rm D}_{\b} ~~~,  ~~~
\D{}^{(165) \, \un{a}\un{b}\un{c}} ~=~ (\g^{\un{a}\un{b}\un{c}})
{}^{\a \b} \, {\rm D}_{\a} \, {\rm D}_{\b}  ~~~, ~~~
\D{}^{(330) \, \un{a}\un{b}\un{c}\un{d}} ~=~ (\g^{\un{a}\un{b}
\un{c}\un{d}}){}^{\a\b} \, {\rm D}_{\a} \, {\rm D}_{\b} ~~~. }  \ee
The implication of the existence of these operators is there is no component
field at quadratic order in ${\cal V}{}^{(11)}$ that occurs in the $\{11\}$ representation
of the 11D Lorentz group.  This should be contrasted with the situation in 
4D superspace where the operator $ (\g{}^5 \g{}_{\un a}){}^{a b} {\rm D}{
}_a  {\rm D}{}_b$ is in the $\{4\}$ representation of the 4D Lorentz group.

However, at quartic order utilizing the 11D spinorial derivatives we can define a superfield 
by the equation
\be
v{}_{\un a}^{(11)} ~=~  \fracm 1{3!} \left[ \D{}^{(165) \, \un{b}\un{c} \un{d}} 
 \D{}^{(330)}{}_{\un{a}\un{b}\un{c}\un{d}} \, {\cal V}{}^{(11)} \right] ~~~,
\ee
that is the analog to one of the equations seen to occur in (\ref{fielddefs}).  The
``Breitenlohner Approach'' can be followed by defining an operator
valued supergravity co-vector $\bm {\rm {SG}}{}_{\un a}$ through the equation
\be  \eqalign{  {~~~~}
{\bm {\rm {SG}}}{}_{\un a} ~=~  &\fracm 1{3!}\left[ \D{}^{(165) \, \un{c}\un{d} \un{e}} 
 \D{}^{(330)}{}_{\un{a}\un{c}\un{d}\un{e}} \, {\cal V}{}^{(11) \, \un{b}} \right] \,
 \pa{}_{\un b} 
 ~+~ \fracm 1{3!}
 \left[ \D{}^{(165) \, \un{b}\un{c} \un{d}} 
 \D{}^{(330)}{}_{\un{a}\un{b}\un{c}\un{d}} \, {\cal V}{}^{(11) \, \b} \right] \,
 {\rm D}{}_{\b} 
 ~+~   \cr
 &\fracm 1{2 \cdot 3!} \left[ \D{}^{(165) \, \un{b}\un{c} \un{d}} 
 \D{}^{(330)}{}_{\un{a}\un{b}\un{c}\un{d}} \, {\cal V}{}^{(11) \, \un{k}\, \un{l}} \right] \,
 {\cal M}{}_{\un{k} \, \un{l} }  ~~~,  \label{SGdef}
}\ee  
and above $ \pa{}_{\un b}$, $ {\rm D}{}_{\b}$, ${\cal M}{}_{\un{k} \, \un{l}}$ denote 
respectively the 11D partial derivative operator, the 11D spinor superspace
derivative, and the 11D Lorentz generators.  There are other difference also.  For 
the 4D superfield $V$ need only to be expanded to quartic order in $\theta{}^a$.
In the case of the ${\cal V}{}^{(11)}$ the $\theta{}^a$-expansion goes out to the
order of $\theta$ raised to the thirty second power.

If no obstructions occur, this will describe 11D SG in superspace just as the expressions 
in (1.1) - (1.4) did for 4D, $\cal N$ = 1 superspace.  It is possible the superfields ${\cal 
V}{}^{(11) \, \un{b}}$, $ {\cal V}{}^{(11) \, \b}$, and ${\cal V}{}^{(11) \, \un{k}\, \un{l}}$ can 
be expressed in terms of more fundamental superfields as is the case in 4D, 
$\cal N$ = 2 superspace supergravity \cite{N2sg}.  Moreover, were 11D superspace 
supergravity to follow the pattern of its lower dimensional ``relatives,'' the conformal 
part of the 11D graviton will be contained in the first term and the conformal part of the 
11D gravitino will be contained in the second term of (\ref{SGdef}).

In any case, this approach would put a maximum limit on the number
of component fields required in ${\bm {\rm {SG}}}{}_{\un a}$.  We simply 
count the number of free indices on ${\cal V}{}^{(11) \, \un{b}}$, $ {\cal V}{}^{(11) \, \b}$, 
and ${\cal V}{}^{(11) \, \un{k}\, \un{l}}$ to find 11 + 32 + 55 = 98 and multiply
by the number of component fields in ${\cal V}{}^{(11)}$ to arrive 
at 210,453,397,504 bosonic component fields and  210,453,397,504 fermionic 
component fields.  In fact depending on the size of the null space (and
thus the gauge transformations of ${\cal V}{}^{(11)}$) of the condition
\be
\left[ \D{}^{(165) \, \un{b}\un{c} \un{d}} 
\D{}^{(330)}{}_{[ \un{a}|\un{b}\un{c}\un{d}} \, \pa{}_{| \un{e} ]}
 \left( \d {\cal V}{}^{(11)} \right) \right] ~=~ 0 ~~~,
\ee
the numbers could even be considerably less.  There is also an argument that 
can be made to estimate the lower bound on the number of component fields 
involved.  

In the works of \cite{GRana1,GRana2} an algorithm was presented that, given a theory
that possesses a number of 1D supercharges, determines the size of the smallest
 {\em {irreducible}} 1D SUSY representation.  The theory in 11D, when reduced 
to 1D, corresponds to a 1D theory with 32 supercharges.  For 1D, $N$ = 32 
supersymmetry the smallest off-shell representations determined by the algorithm
possess 32,768 bosonic fields and 32,768 fermionic fields.  Once more multiplying by 
98 we are led to a lower bound of 3,211,264 bosonic components and 3,211,264 
fermionic components.  While 3.2 million component fields may seem a large number, 
it is far less than one percent of 210 billion.

Perhaps now the stage is set for us to (and very roughly paraphrasing Hilbert -
reach beyond the level of G\" ottingen's children) understand off-shell eleven 
dimensional supergravity supergeometry for M-Theory... and as well (with appropriate
modifications), for ten dimensional supergravity supergeometries for heterotic and 
superstrings.

 \vspace{.05in}
 \begin{center}
\parbox{4in}{{\it ``Every boy in the streets of G\" ottingen understands 
more $~$ about  four dimensional geometry than Einstein.  Yet, in 
$~$ spite of that, 
Einstein did the work and not the mathe- $~$ maticians.'' \\ ${~}$ 
\\ ${~}$ }\,\,-\,\, David Hilbert $~~~~~~~~~$}
 \parbox{4in}{
 $~~$}  
 \end{center}
$$~~$$  
 \noindent
{\bf Acknowledgements}\\[.1in] \indent
We would like to recongnize Martin Cederwall, William Linch, and Warren Siegel for
conversations.  The research of S.\ J.\ Gates, Jr., Y.\ Hu, and S.-N.\ Mak is supported by 
the endowment of the Ford Foundation Professorship of Physics at Brown University and 
partially supported by the U.S. National Science Foundation grant PHY-1315155.

\newpage
$$~~$$


\begin{thebibliography}{99}
\small\frenchspacing\raggedright

\bibitem{N1}
G.\ Nordstr\" om, ``Inertial and gravitational mass in relativistic mechanics,'' Annalen 
der Physik, {\bf {40}}  856 (1913).

\bibitem{N2}
G.\ Nordstr\" om, ``About the theory of gravitation from the point of view of the 
principle of relativity,'' Annalen der Physik, {\bf {42}}  533 (1913).

\bibitem{CST1}
H.\ Nishino, and S.\ J.\ Gates, Jr., 
``Chern-Simons Theories with Supersymmetries in Three Dimensions,''
Int.\ J.\ Mod.\ Phys.\ {\bf {A8}} (1993) 3371-3422,
DOI: 10.1142/S0217751X93001363. 

\bibitem{CST2}
 S.\ J.\ Gates, Jr., and H.\ Nishino, 
 ``Remarks on N= 2 Supersymmetric Chern-Simons Theories,''
 Phys.\ Lett.\ {\bf { B281}} (1992) 72-80
DOI: 10.1016/0370-2693(92)90277-B. 

\bibitem{CST3}
O.\ Aharony, O.\ Bergman, Danie L.\ Jafferis, and J.\ Maldacena,
``N=6 Superconformal Chern-Simons-Matter Theories, M2-branes 
and Their Gravity Duals,''
JHEP {\bf {0810}} (2008) 091,
DOI: 10.1088/1126-6708/2008/10/091, e-Print: arXiv:0806.1218 [hep-th].

\bibitem{PF1}
S.\ M.\ Chester, S.\ S.\ Pufu, and X.\ Yin,
``The M-Theory S-Matrix From ABJM: Beyond 11D Supergravity,''
JHEP {\bf {1808}} (2018) 115,
DOI: 10.1007/JHEP08(2018)115,
e-Print: arXiv:1804.00949 [hep-th].

\bibitem{PF2}
D.\ J.\ Binder, S.\ M.\ Chester, and S.\ S.\ Pufu, 
``Absence of D${}^4$ R${}^4$ in M-Theory From ABJM,''
PUPT-2570,
e-Print: arXiv:1808.10554 [hep-th].

\bibitem{NordSG1}
S.\ J.\ Gates, Y.\ Hu, H.\ Jiang, and S.-N.\ Hazel Mak,
``On Linearized Nordstr\"om Supergravity in Eleven and Ten Dimensional Superspaces,'' 
Brown Univ. HET-1779
e-Print: arXiv:1812.05097 [hep-th].

\bibitem{Ssp8c}
A.\  Salam, and J.\ A.\ Strathdee,
``On Superfields and Fermi-Bose Symmetry,''
Phys.\ Rev.\ {\bf {D11}} (1975) 152,
DOI: 10.1103/PhysRevD.11.1521 

\bibitem{WZ1}
J.\ Wess, and B.\ Zumino, ``Superspace Formulation of Supergravity,''
Phys.\ Lett.\ {\bf {66B}} (1977) 361, DOI: 10.1016/0370-2693(77)90015-6.

\bibitem{WZ2}
J.\ Wess, and B.\ Zumino, ``Superfield Lagrangian for Supergravity,''
Phys.\ Lett.\ {\bf {74B}} (1978) 51, DOI:  10.1016/0370-2693(78)90057-6.

\bibitem{P1}
B.\ E.\ W.\ Nilsson, and A.\ K.\ Tollsten,	
``The Geometrical Off-shell Structure of Pure N=1, d=10 Supergravity in Superspace,''
Phys.\ Lett.\ {\bf {169B}} (1986) 369, DOI: 10.1016/0370-2693(86)90374-6. 

\bibitem{P2}
P.\ S.\ Howe, ``Pure spinor lines in superspace and ten-dimensional supersymmetric 
theories'', Phys.\ Lett.\ {\bf {B258}} (1991) 141, Addendum: Phys.\ Lett.\ {\bf {B259}} 
(1991) 511. DOI: 10.1016/0370-2693(91)91221-G.

\bibitem{P3}
P.\ S.\ Howe, ``Pure spinors, function superspaces and supergravity theories in ten 
and eleven dimensions'', Phys.\ Lett.\ {bf {B273}} (1991) 90, DOI: 10.1016/0370-2693(91)90558-8.

\bibitem{P4}
M.\ Cederwall, U.\ Gran, M.\ Nielsen and B.\ E.\ W. Nilsson, ``Manifestly supersymmetric 
M-theory'', J.High Energy Phys. 0010 (2000) 041 [hep-th/0007035]; ``Generalised 
11-dimensional supergravity'', hepth/0010042, DOI: 10.1088/1126-6708/2000/10/041.

\bibitem{P5}
P.\ S.\ Howe, and N.\ Berkovits, ``Ten-dimensional supergravity constraints from the pure spinor 
formalism for the superstring,'' Nucl.\ Phys.\ {\bf {B635}} (2002) 75, e-Print: hep-th/0112160
DOI: 10.1016/S0550-3213(02)00352-8.
 	
\bibitem{P6}	
M.\ Cederwall, ``D=11 supergravity with manifest supersymmetry,'' Mod.\ Phys.\
Lett.\ {\bf {A25}} (2010) 3201, e-Print: arXiv:1001.0112 [hep-th],
DOI: 10.1142/S0217732310034407.

\bibitem{P7}
M.\ Cederwall, ``Towards a manifestly supersymmetric action for 11-dimensional supergravity,''
JHEP {\bf {1001}} (2010) 117, e-Print: arXiv:0912.1814 [hep-th], DOI: 10.1007/JHEP01(2010)117.

\bibitem{Q1}
O.\ Bedoya, and N.\ Berkovits,
``GGI Lectures on the Pure Spinor Formalism of the Superstring,''
Presented at Conference: C09-04-06.1
e-Print: arXiv:0910.2254 [hep-th].

\bibitem{Q2}
H.\ Gomez, and C.\ Mafra,
``The closed-string 3-loop amplitude and S-duality,''
JHEP {\bf {1310}} (2013) 217, e-Print: arXiv:1308.6567 [hep-th], 
DOI: 10.1007/JHEP10(2013)217.

\bibitem{Q3}
N.\ Berkovits, and N.\ Nekrasov, 
``Multiloop Superstring Amplitudes from Non-Minimal Pure Spinor Formalism,''
JHEP {\bf {0612}} 029,2006, e-print: arXiv:hep-th/0609012,
DOI: 	10.1088/1126-6708/2006/12/029.

\bibitem{Q4}
N.\  Berkovits, and E.\  Witten, 
``Supersymmetry Breaking Effects using the Pure Spinor Formalism of the Superstring,''
JHEP {\bf {1406}} (2014) 127, e-Print: arXiv:1404.5346 [hep-th], 
DOI: 10.1007/JHEP06(2014)127.

\bibitem{B1}
P.\ Breitenlohner, ``A Geometric Interpretation of Local Supersymmetry,''
Phys.\ Lett.\ {\bf {67B}} (1977) 49, DOI: 10.1016/0370-2693(77)90802-4.

\bibitem{SpRSp8BK}
S.\ J.\ Gates Jr., M.\ T.\ Grisaru, M.\ Roc\v ek, and W.\ Siegel, 
``Superspace Or One Thousand and One Lessons in Supersymmetry,''
Front.\ Phys.\ {\bf {58}} (1983) 1, e-Print: hep-th/0108200.

\bibitem{S1}
W.\ Siegel,
``Supergravity Superfields Without a Supermetric,'' (Harvard U.). Nov 1977. 16 pp., Harvard
Univ. preprint, HUTP-77/A068, (unpublished). 

\bibitem{SFSG}
W.\ Siegel, S.\ J.\ Gates, Jr.,
``Superfield Supergravity,'' 
Nucl.\ Phys.\ {\bf {B147}} (1979) 77-104,
DOI: 10.1016/0550-3213(79)90416-4. 

\bibitem{2MT}
H.\ Nishino, and S.\ J.\ Gates, Jr., 
``Toward an Off - Shell 11D Supergravity Limit of M - Theory,''
Phys.\ Lett.\ {\bf {B388}} (1996) 504-511,
DOI: 10.1016/S0370-2693(96)01193-8,
e-Print: hep-th/9602011.

\bibitem{2MTcf}
S.\ J.\ Gates, Jr., 
``Superconformal symmetry in 11-D superspace and the M theory effective action,''
Nucl.\ Phys.\ {\bf {B616}} (2001) 85-105,
DOI: 10.1016/S0550-3213(01)00421-7.

\bibitem{N2sg}
S.\ J.\ Gates Jr.,  and W.\ Siegel, 
``Linearized N=2 Superfield Supergravity,''
Nucl.\ Phys.\ {\bf {B195}} (1982) 39-60;
DOI: 10.1016/0550-3213(82)90047-5. 

\bibitem{GRana1}
S.\ J.\ Gates, Jr., and L.\ Rana,  ``A Theory of Spinning Particles for Large 
N-extended Supersymmetry (I),'' Phys.\ Lett.\ {\bf {B352}} (1995) 50;
DOI: 10.1016/0370-2693(95)00474-Y, e-Print: arXiv [hep-th:9504025].

\bibitem{GRana2}
S.\ J.\ Gates Jr., and L.\ Rana, ``A Theory of Spinning 
Particles for Large N-extended Supersymmetry (II),'' ibid.\ Phys.\ Lett.\ {\bf 
{B369}} (1996) 262; DOI: 10.1016/0370-2693(95)01542-6, e-Print: arXiv [hep-th:9510151].

\end{thebibliography}
\end{document}